\documentclass[fleqn, usenatbib]{mnras}
\usepackage{newtxtext,newtxmath}
\usepackage[T1]{fontenc}

\DeclareRobustCommand{\VAN}[3]{#2}
\let\VANthebibliography\thebibliography
\def\thebibliography{\DeclareRobustCommand{\VAN}[3]{##3}\VANthebibliography}

\usepackage{graphicx}	% Including figure files
\usepackage{amsmath}	% Advanced maths commands
 
\title[Stripped giants in the binary zoo]{Unicorns and Giraffes in the binary zoo: stripped giants with subgiant companions}

\author[El-Badry et al.]{Kareem El-Badry,$^{1,2,3}$\thanks{E-mail: kareem.el-badry@cfa.harvard.edu} 
Rhys Seeburger,$^3$
Tharindu Jayasinghe,$^4$
Hans-Walter Rix,$^3$
Silvia Almada,$^3$  \newauthor
Charlie Conroy,$^1$ 
Adrian M. Price-Whelan,$^5$ 
Kevin Burdge$^{6,7}$  \\
$^{1}$Center for Astrophysics $|$ Harvard \& Smithsonian, 60 Garden Street, Cambridge, MA 02138, USA\\
$^{2}$Harvard Society of Fellows, 78 Mount Auburn Street, Cambridge, MA 02138\\
$^{3}$Max-Planck Institute for Astronomy, K\"onigstuhl 17, D-69117 Heidelberg, Germany\\
$^{4}$Department of Astronomy, The Ohio State University, 140 West 18th Avenue, Columbus, OH 43210, USA\\
$^{5}$Center for Computational Astrophysics, Flatiron Institute, 162 Fifth Avenue, New York, NY 10010, USA\\
$^{6}$Department of Physics, Massachusetts Institute of Technology, Cambridge, MA 02139, USA\\
$^{7}$Kavli Institute for Astrophysics and Space Research, Massachusetts Institute of Technology, Cambridge, MA 02139, USA \\
}

\date{\vspace{-1.0cm}}

\pubyear{2022}

\begin{document}
\label{firstpage}
\pagerange{\pageref{firstpage}--\pageref{lastpage}}
\maketitle

\begin{abstract}
We analyze two binary systems containing giant stars, V723 Mon (``the Unicorn'') and 2M04123153+6738486 (``the Giraffe''). Both giants orbit more massive but less luminous companions, previously proposed to be mass-gap black holes. Spectral disentangling reveals luminous companions with star-like spectra in both systems. Joint modeling of the spectra, light curves, and spectral energy distributions robustly constrains the masses, temperatures, and radii of both components: the primaries are luminous, cool giants ($T_{\rm eff,\,giant} = 3,800\,\rm K$ and $4,000\,\rm K$, $R_{\rm giant}= 22.5\,R_{\odot}$ and $25\,R_{\odot}$) with exceptionally low masses ($M_{\rm giant} \approx 0.4\,M_{\odot}$) that likely fill their Roche lobes. The secondaries are only slightly warmer subgiants ($T_{\rm eff,\,2} = 5,800\,\rm K$ and $5,150\,\rm K$, $R_2= 8.3\,R_{\odot}$ and $9\,R_{\odot}$) and thus are consistent with observed UV limits that would rule out main-sequence stars with similar masses ($M_2 \approx 2.8\,M_{\odot}$ and $\approx 1.8\,M_{\odot}$). In the Unicorn, rapid rotation blurs the spectral lines of the subgiant, making it challenging to detect even at wavelengths where it dominates the total light.  Both giants have surface abundances indicative of CNO processing and subsequent envelope stripping. The properties of both systems can be reproduced by binary evolution models in which a $1-2\,M_{\odot}$ primary is stripped by a companion as it ascends the giant branch. The fact that the companions are also evolved implies either that the initial mass ratio was very near unity, or that the companions are temporarily inflated due to rapid accretion. The Unicorn and Giraffe offer a window into into a rarely-observed phase of binary evolution preceding the formation of wide-orbit helium white dwarfs, and eventually, compact binaries containing two helium white dwarfs.
\end{abstract}

\begin{keywords}
stars: black holes -- binaries: spectroscopic -- stars: evolution -- stars: individual: V723 Mon -- stars: individual: 2MASS J04123153+6738486 
\vspace{-0.5cm}
\end{keywords}

\section{Introduction}

Identification of detached stellar-mass black holes (BHs) in binaries is a major goal of numerous spectroscopic and photometric surveys. Observational campaigns have have been ongoing for more than half a century \citep[e.g.][]{Guseinov1966, Trimble1969}, but the number of published candidates has increased dramatically in the last few years \citep[e.g.][]{Thompson2019, Liu2019, Giesers2019, Rivinius2020,Gomez2021, Lennon2021, Saracino2022}, spurred by the proliferation of wide-field surveys and by growing interest in the progenitors of gravitational wave sources. Most candidates identified to date are controversial, and non-BH explanations for the observed data have been proposed for many systems \citep[e.g.][]{vandenHeuvel2020, Shenar2020, Bodensteiner2020, El-Badry2021hr6819, El-Badry2022, Stevance2022, El-Badry2022_2004, Frost2022}. In several of these systems, spectral disentangling has shown the presence of two luminous stars directly.

Two recently identified candidates are V723 Mon \citep[``the Unicorn'';][]{Jayasinghe2021} and 2M04123153+6738486 \citep[``the Giraffe'';][]{Jayasinghe2022}. Both systems contain red giant stars in relatively wide orbits (60 and 81 days) and exhibit strong ellipsoidal variability due to tidal deformation of the giants. The giants' orbits imply more massive companions: $M_2 \approx 1.5-3.5\,M_{\odot}$, depending on the assumed inclination and giant mass. But both systems are faint in the UV, ruling out main-sequence companions of the required mass.

Although normal, main-sequence companions are ruled out, there is some evidence of a second luminous source in both systems. In the Unicorn, this is manifest as apparent continuum dilution of the giant's absorption lines at blue wavelengths; in the Giraffe, there is evidence of a second set of absorption lines. Phase-dependent variability of the Balmer lines is observed in both systems, suggesting ongoing mass transfer. Previous analyses \citep{Jayasinghe2021, Jayasinghe2022} concluded that the second luminous source in both systems is too faint to be a stellar companion of the required mass. These works proposed that the second light originates from an accretion disk around a BH companion (in the Giraffe), or from a diffuse circumbinary nebula (in the Unicorn).

There is no robust X-ray detection in either system. If the companions are BHs that accrete the giants' winds at the Hoyle-Lyttleton rate, the observed upper limits ($L_{\rm X} \lesssim 10^{30}\,\rm erg\,s^{-1}$ and $L_{\rm X} \lesssim 10^{32}\,\rm erg\,s^{-1}$) imply radiative efficiencies of $\eta \lesssim 10^{-5}$. If the second light is interpreted as arising from accretion, the accretion flows would have X-ray-to-optical luminosity ratios $L_X/L_{\rm opt} \lesssim 10^{-3}$. 

In this paper, we study the luminous secondaries in both systems in more detail, using a combination of spectral disentangling and fitting of binary spectral models. Given that the optical and X-ray properties of the secondaries are different from what is expected from an accretion flow onto a BH, we ask whether there is a plausible scenario in which the observed data can be explained by two luminous stars. We find that such a scenario does exist in both systems and reproduces the observed spectra, spectral energy distributions, and light curves more successfully than scenarios involving a BH. 

The rest of this paper is organized as follows. Section~\ref{sec:summary_both} summarizes previous analyses of both systems. In Section~\ref{sec:data}, we reanalyze the available data for the Giraffe, including the optical and infrared spectra, broadband spectral energy distribution (SED), and light curves. In Section~\ref{sec:data_unicorn}, we carry out a similar analysis for the Unicorn. We investigate the formation history and future evolution of the systems using binary evolution calculations  in Section~\ref{sec:evol}. We summarize our results and discuss the systems in the context the broader binary population in Section~\ref{sec:discussion}.

\section{Summary of both systems}
\label{sec:summary_both}

\subsection{The Giraffe}
\label{sec:giraffe_summary}
The Giraffe \citep{Jayasinghe2022} is a binary in Camelopardalis containing a luminous red giant ($L_{\rm giant}\approx 200\,L_{\odot}$) and a companion in an 81-day circular orbit. The spectroscopic mass function, $f(M_2)\equiv \bigl ( M_2\sin{i} \bigr )^3/(M_{\rm giant}+M_2)^2 = 0.58\,M_{\odot}$, implies a relatively massive companion. By modeling the system's SED, light curves, and RVs, \citet{Jayasinghe2022} inferred a giant mass of $\approx 0.6\,M_{\odot}$, inclination of $\approx 41$ degrees, and unseen companion mass of $\approx 3\,M_{\odot}$.\footnote{The \citet{Jayasinghe2022} model for the Giraffe to which we refer in this paper corresponds to the preprint dated January 28, 2022. As this is a preprint, the final version of that paper may ultimately present different models.} A main-sequence star of this mass would overwhelm the observed SED at blue wavelengths, so they concluded that the companion is a BH.   

When they subtracted the best-fit spectral model for the giant from the observed optical spectra, \citet{Jayasinghe2022} found residuals resembling the narrow-lined absorption spectrum of a star. The residual spectrum moves in anti-phase with the giant, but with an RV amplitude 5 times smaller. Assuming these RVs trace the reflex motion of the companion's center of mass, they imply a mass ratio $M_{\rm giant}/M_{2} \approx 0.2$. Ascribing this secondary spectrum to a stellar companion was noted as the most natural explanation by \citealt{Jayasinghe2022}). But they found that a normal star of the required mass would produce a UV flux inconsistent with the observed limits. This led to their proposal that the secondary spectrum traces an accretion disk around a BH companion. 

No X-rays were detected, and the observed upper limit is $L_{X}\lesssim 10^{32}\,\rm erg\,s^{-1}$. Relatively strong, double-peaked H$\alpha$ emission was observed, suggesting that some sort of mass transfer is likely ongoing. \citet{Jayasinghe2022} found that the giant does not fill its Roche lobe but nearly does, with $R/R_{\rm Roche\,lobe}\approx 0.93$. This, and the inferred inclination, is sensitive to the amount of dilution from the secondary. The distance to the system is of order $3-4$\,kpc; as we discuss in Section~\ref{sec:distance_giraffe}, the exact value is sensitive to the {\it Gaia} parallax zeropoint and assumptions about the nature of the system.

\subsection{The Unicorn}
\label{sec:unicorn_summary}
V723 Mon (``the Unicorn'') is a bright, nearby giant in Monoceros. It was identified as a 60-day spectroscopic binary by \citet{Griffin2010}, with the orbit refined by \citet{Strassmeier2012} and \citet{Griffin2014}. All works found an orbital period $P_{\rm orb} \approx 60$ days, RV semi-amplitude for the giant $K_{\rm giant}\approx 65\,\rm km\,s^{-1}$, and an eccentricity near 0, yielding a mass function $f(M_2)\approx 1.71\,M_{\odot}$. \citet{Strassmeier2012} found small ($\approx 1-2\,\rm km\,s^{-1}$) but significant residuals when fitting the giant's RVs with a Keplerian orbit. They also found evidence of a luminous secondary in their spectra, which moved approximately in anti-phase with the giant. The RVs of this secondary component were poorly fit with a Keplerian orbit. \citet{Strassmeier2012} concluded that the system is a hierarchical triple, with a period of $\approx 20$ days for the inner binary and a total inner binary mass of 3-4 times the giant mass.

V723 Mon was re-analyzed by \citet{Jayasinghe2021}, who argued that the system contains a BH. They found that the triple model proposed by \citet{Strassmeier2012} was not viable because (a) the system would not be dynamically stable and (b) variations in the tidal force on the giant due to the orbit of the inner binary would make the outer orbit eccentric and lead to observable variations in its tidal deformation. \citet{Jayasinghe2021} also noted that tidal deformation of the giant is expected to cause changes in the giant's line profiles with orbital phase. They proposed that this tidal deformation is responsible for both the apparent deviations of the giant's RVs from a Keplerian orbit (as explored quantitatively by \citealt{Masuda2021}) and for the apparent evidence of a luminous secondary in the spectra. 

\citet{Jayasinghe2021} found that the giant's absorption lines are shallower than expected at blue wavelengths, which they interpreted as ``veiling'' by another light source. The ellipsoidal variability amplitude also becomes weaker toward bluer wavelengths, suggesting dilution by a non-variable object that is warmer than the giant. They did not find the spectrum of this diluting component to have absorption lines or other star-like features and therefore attributed it to continuum from a diffuse circumbinary component. 

The depth and width of the Unicorn's Balmer lines varies with orbital phase, with the lines appearing deepest at phase $\phi \approx 0.5$. \citet{Jayasinghe2021} attributed this to emission from an accretion disk surrounding the putative BH, which would be eclipsed at phase $\phi =0.5$. Weak X-rays were detected in {\it Swift} XRT data, with a luminosity $L_{X}\sim 10^{30}\,\rm erg\,s^{-1}$, but \citet{Hare2021} showed that this detection is likely spurious. The X-ray luminosity is less than $10^{-5} \times$ the inferred optical luminosity of the veiling component. 

After correcting for the {\it Gaia} parallax zeropoint \citep[][]{Lindegren2021} and underestimated parallax uncertainties, the inferred parallax of V723 Mon is $2.20\pm 0.04$ mas, corresponding to a distance of $454\pm 8$ pc. Because the system is nearby, its inferred properties are only weakly sensitive to assumptions about the parallax zeropoint.

\section{Data and modeling: the Giraffe}
\label{sec:data}

We begin with the Giraffe. The narrow lines of the secondary in this system make the analysis of the spectra more straightforward than in the Unicorn.

\subsection{Near-infrared spectra}
\label{sec:apogee_summary}

The Giraffe was observed by the APOGEE survey \citep[][]{Majewski2017}, which obtained spectra at 4 epochs with resolution $R\approx 22,500$ in the $H-$band (15,000-17,000\,\AA). We fit these spectra with a binary spectral model, as described in detail in Appendix~\ref{sec:apogee}. 

We find that two luminous components contribute to the APOGEE spectra: a giant with effective temperature of $4050\pm 100$\,K, and a subgiant with  effective temperature $5200\pm 200$. The giant dominates the $H-$band light, contributing 80\% of the flux, but the 20\% contribution from the secondary component, appearing to be a subgiant, is unambiguous. This is somewhat unexpected, because \citet{Jayasinghe2022} found that the light contributions of the secondary detected in the optical became negligible at $\lambda \gtrsim 7,000$\,\AA. We find the subgiant to move in anti-phase with the giant, with an implied dynamical mass ratio $M_{\rm giant}/M_{2} =0.21\pm 0.02$. This is in good agreement with our findings from the optical spectra and implies that we detect the same secondary in the near-infrared and optical, and derive consistent characteristics.

\subsection{Optical spectra}
\label{sec:HIRES}

\begin{figure*}
    \centering
    \includegraphics[width=\textwidth]{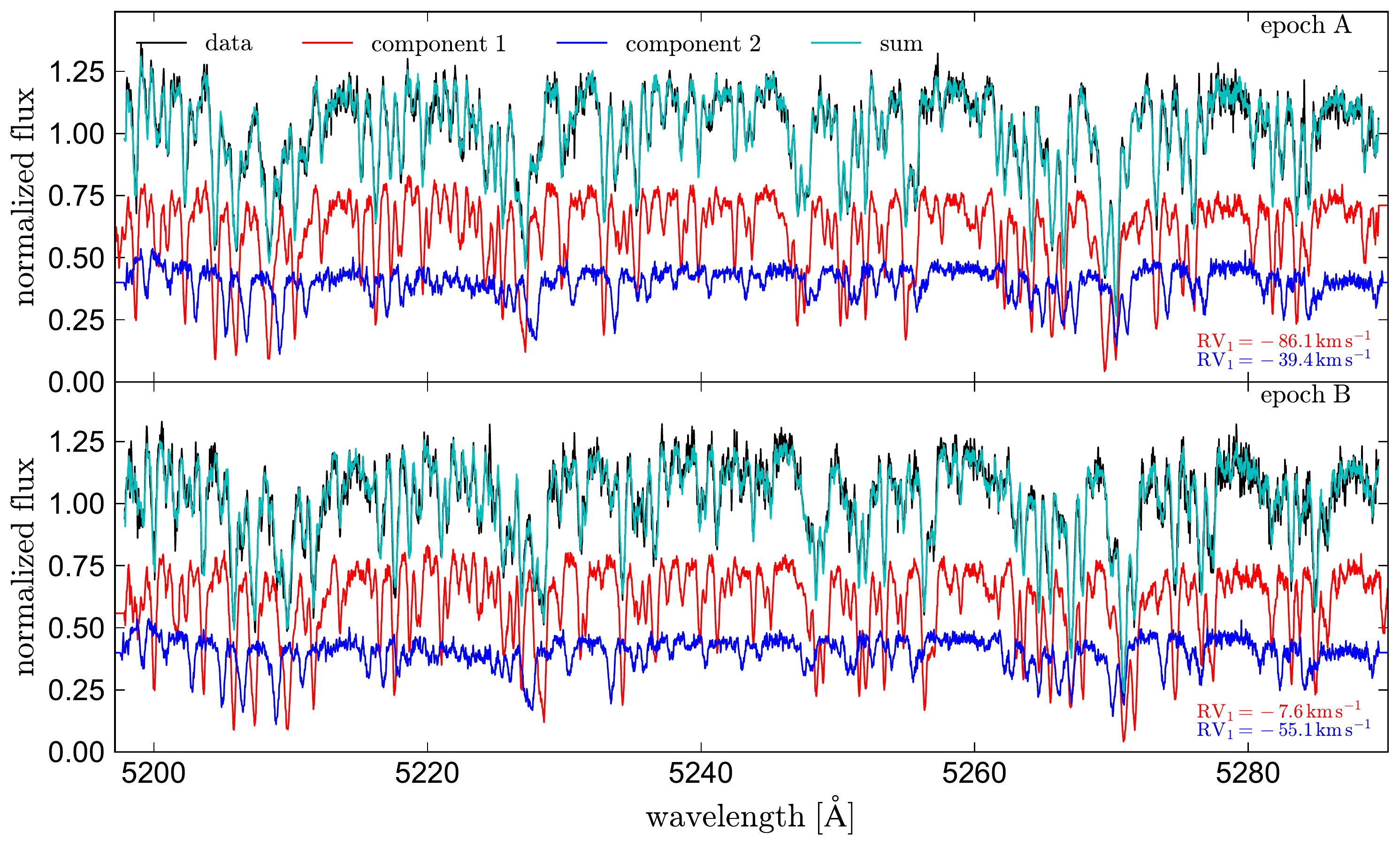}
    \caption{Spectral disentangling of the Giraffe system. Top and bottom panels show two different epochs (among a total of 8), chosen to be near opposite quadratures. Black lines show the observed Keck/HIRES spectra. Blue and red lines show the disentangled spectra of the giant and subgiant, which are identical in both panels except for RV shifts and whose sum (cyan) matches the observed spectra, as well as those in the other 6 epochs that are not shown. Both components resemble stellar spectra (see Figure~\ref{fig:giraffe_models}). }
    \label{fig:giraffe_disentangle}
\end{figure*}

We analyzed 8 epochs of Keck/HIRES spectra of the Giraffe, which are sampled roughly uniformly in orbital phase and are described in detail in \citet{Jayasinghe2022}. These data provide useful coverage of most of the wavelength range between 3900 and 8000 \AA, with resolution $R\approx 60,000$ and typical per-epoch SNR of 20 per pixel at 5000\,\AA.

We first fit the optical spectra using a binary spectral model, following the approach outlined in Appendix~\ref{sec:apogee} and \citet{El-Badry2018}. We fit all usable orders simultaneously, self-consistently accounting for the wavelength-dependent flux ratio. We fit separate visits independently, estimating uncertainties based on the epoch-to-epoch scatter. The spectral model was trained on synthetic spectra from the BOSZ grid \citep[][]{Bohlin2017} calculated with ATLAS12 and SYNTHE \citep[][]{Kurucz_1970, Kurucz_1979, Kurucz_1993}. In good agreement with the results from the APOGEE spectra, we found a giant with effective temperature of $4000\pm 100$\,K, and a subgiant with effective temperature $5150\pm 200$\,K. The best-fit metallicity (assumed the same for both components) is $[\rm Fe/H] \approx -0.5$. The subgiant's inferred flux contributions increase toward the blue orders: we find that it contributes $\approx 35\%$ of the light at 6,000\,\AA, but $\approx 70\%$ at 4,000\,\AA.

\subsubsection{Spectral disentangling}

\begin{figure*}
    \centering
    \includegraphics[width=\textwidth]{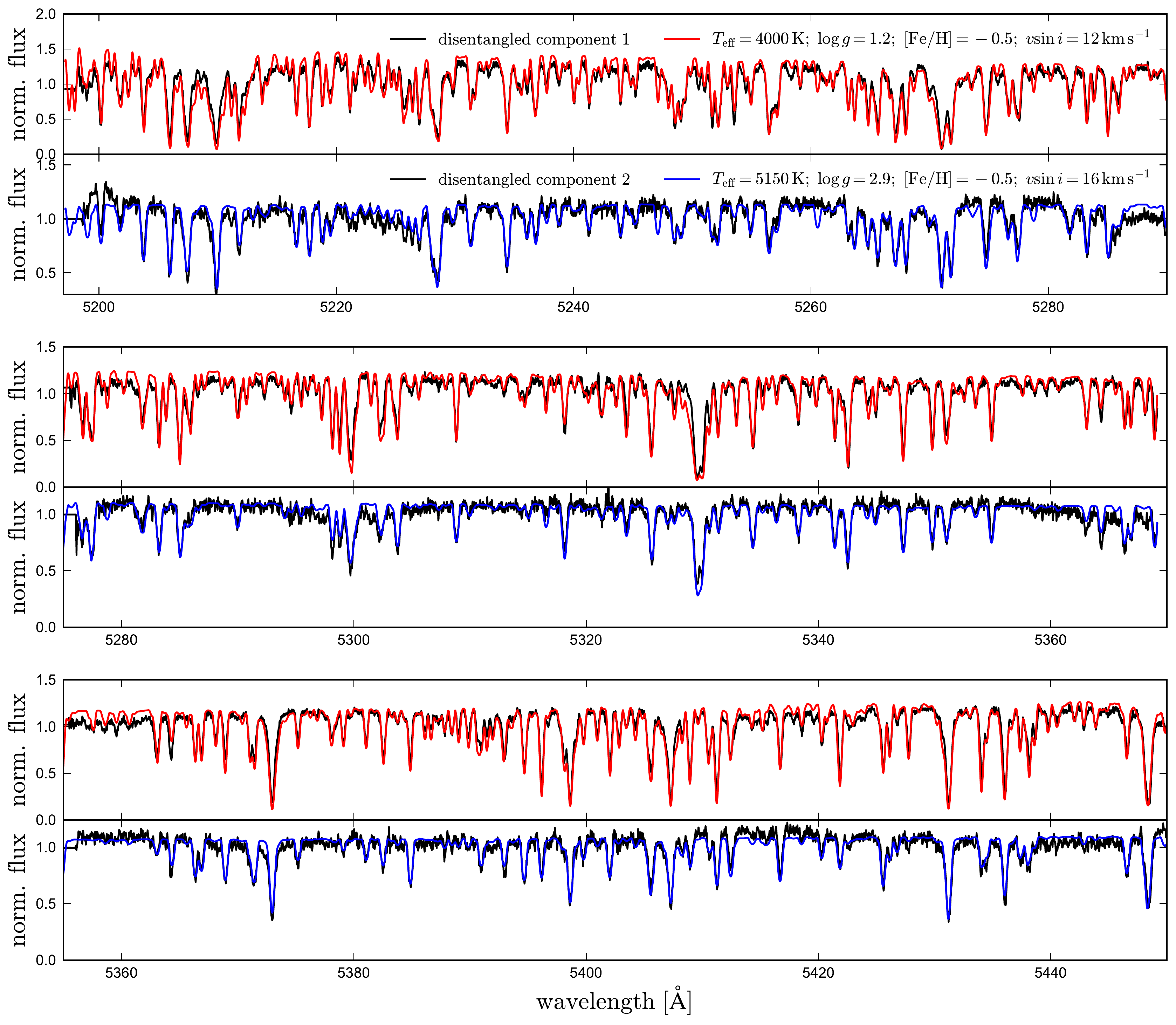}
    \caption{Comparison of disentangled spectra of the Giraffe to spectral models whose stellar labels are noted in the top panel. Black lines show the normalized disentangled spectra of both components; red and blue lines show Kurucz model spectra for a giant and subgiant. Three sets of panels show three different spectral regions. There is qualitatively very good agreement between data and models.  }
    \label{fig:giraffe_models}
\end{figure*}

Our spectral disentangling approach is based on the algorithm presented by \citet{Simon1994}, which aims to solve for the rest-frame component spectra that can best reproduce a set of observed composite spectra under the {\it ansatz} that the component spectra do not change with time and only shift back and forth in RV. Under this assumption, solving for the two component spectra is an overconstrained linear algebra problem, which can be solved without spectral models. We use an optimized implementation of this algorithm, with the addition of a regularization term to minimize spurious undulations introduced in disentangling \citep[e.g.][]{Hensberge2008}, that will be detailed in future work (Seeburger et al., in prep).  

We disentangle several $\approx 100$\,\AA-wide HIRES orders independently. For an initial estimate of the RVs, we use values determined by fitting a binary model. We then use an MCMC sampler \citep{emcee2013} to refine the RVs of the giant and the dynamical mass ratio, under the assumption that the components move in anti-phase and obey conservation of linear momentum. Disentangling does not uniquely determine the flux ratio: there is a perfect degeneracy between the flux ratio and relative line depths \citep[e.g.][]{Ilijic2004}. We set the flux ratio in each order to the median value inferred for that order when fitting the composite spectra with a binary model. Prior to disentangling, we normalize each order by dividing the reported flux by a first order polynomial fit. This is not meant to correspond to the true continuum, which is ill-defined for cool stars in the blue orders, but simply brings all spectra to a common scale. 

Figure~\ref{fig:giraffe_disentangle} illustrates the results of disentangling. We fit all 8 epochs simultaneously but show only two. Disentangling is successful in the sense that all the observed spectra can be well-described as a sum of two Doppler-shifted components. It is clear that two luminous components with absorption spectra are required, because the shape of the composite spectra changes from epoch to epoch. This is also evident from the APOGEE spectra (Appendix~\ref{sec:apogee}) but is more obvious in the optical, where the subgiant contributes a larger fraction of the total light. Our MCMC fitting yields a dynamical mass ratio $M_{\rm giant}/M_{2}=0.198\pm 0.005$. This constraint is robust across independent orders and in good agreement with the value inferred by \citet{Jayasinghe2022}. The same is true for our inferred single-epoch RVs.

In Figure~\ref{fig:giraffe_models}, we compare the disentangled spectra to the best-fit Kurucz models. Consistent with the results from the APOGEE spectra, we find a giant with $T_{\rm eff}\approx 4000\,\rm K$ and a subgiant with $T_{\rm eff}\approx 5150\,\rm K$ (Table~\ref{tab:summary}). Both components have metallicity $\rm [Fe/H] \approx -0.5$ and relatively low $v\sin i$ of $\approx 12$ and $\approx 16\,\rm km\,s^{-1}$. Both components are well-fit by the synthetic spectral models. We emphasize that this does not ``have to'' occur: disentangling imposes no model-based priors on the shape of the component spectra. The fact that both components look like stellar spectra thus increases our confidence that there are two stellar components and that the assumptions built into disentangling are not too far from reality.

\subsubsection{CNO abundances}
Our initial fitting assumed that all metals trace the solar abundance pattern up to a global metallicity scaling factor. To explore possible deviations from this assumption due to CNO processing, we fit the abundances of C, N, and O independently, under the assumption that both components have the same surface abundances. We found $\rm [C/Fe]=-0.5\pm 0.1$,  $\rm [N/Fe]=0.6\pm 0.1$, and $\rm [O/Fe]=0.0\pm 0.1$; that is, a significant enhancement of nitrogen and deficit of carbon compared to the solar abundance pattern. The implied carbon-to-nitrogen ratio, $[\rm C/N]=-1.1$, is extremely low compared to typical low-mass giants \citep[e.g.][]{Hasselquist2019} and strongly suggestive of CNO processing \citep[e.g.][]{El-Badry2022hd15124}.

\subsection{Distance and radius}
\label{sec:distance_giraffe}

\begin{figure*}
    \centering
    \includegraphics[width=\textwidth]{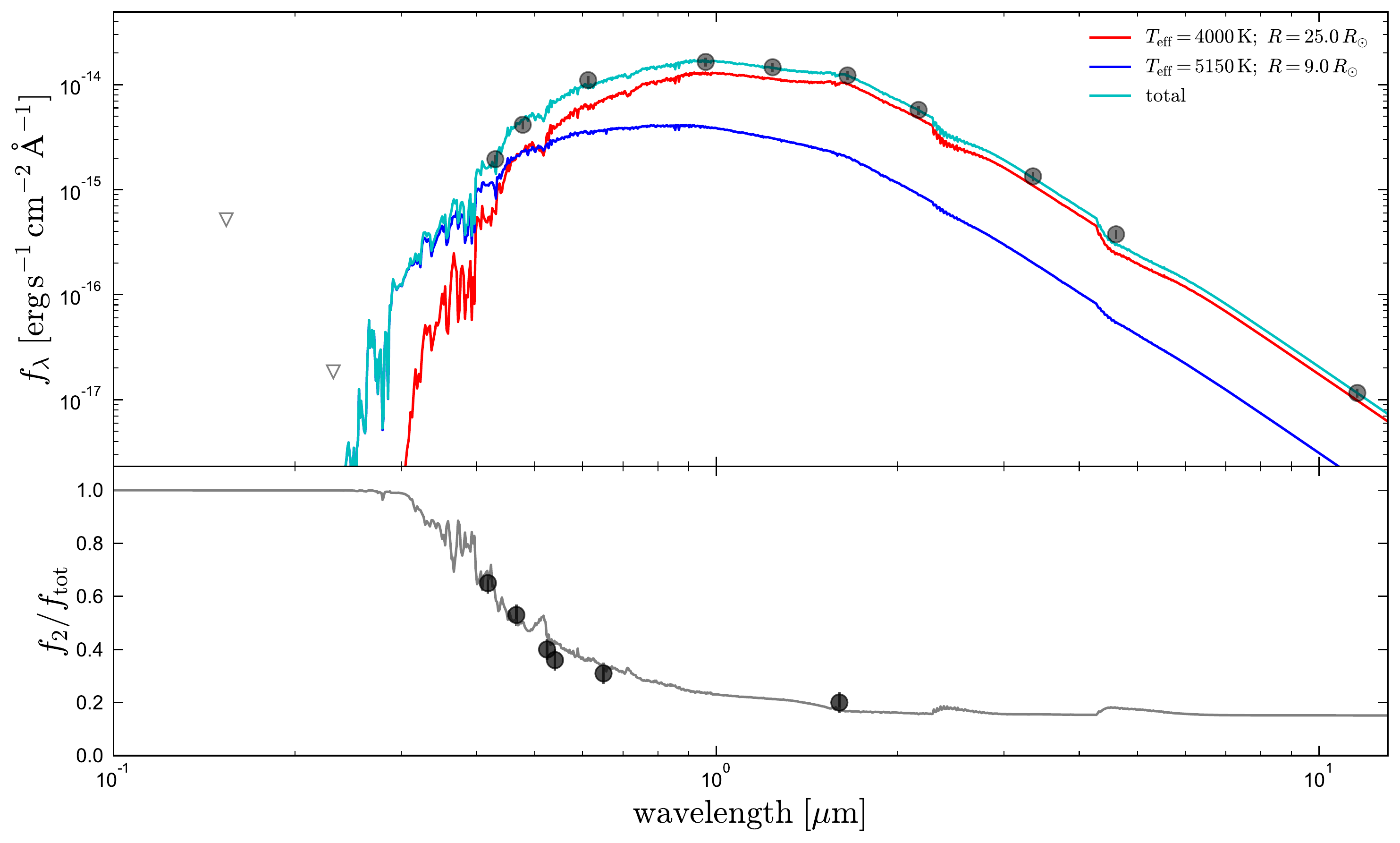}
    \caption{Spectral energy distribution of the Giraffe. Gray points show observed fluxes; inverted triangles show upper limits. We assume a distance of 3.69 kpc and reddening $E(B-V)=0.55$.  Bottom panel shows the fraction of the total light that is contributed by the subgiant (blue in the top panel); gray points show measurements of this quantity from spectral fitting. The secondary is a $\approx 1.7\,M_{\odot}$ subgiant, which contributes about 20\% of the light in the $H$ band,  35\% in the $r$-band, and $\gtrsim 50\%$ at $\lambda \lesssim 5,000$\,\AA. }
    \label{fig:sed}
\end{figure*}
\subsubsection{Distance}
The Giraffe's {\it Gaia} eDR3 parallax  is $\varpi = 0.231 \pm 0.013$. Applying the parallax zeropoint correction from \citet{Lindegren2021} and inflating the parallax uncertainty according to \citet{El-Badry2021_gaia} yields a corrected parallax $\varpi_{\rm corr} = 0.271 \pm 0.017$, corresponding to a distance $d\approx\ 3.69\pm 0.22\,\rm kpc$. The is slightly smaller than the fiducial value of 4 kpc adopted by \cite{Jayasinghe2022}.

\cite{Jayasinghe2022} noted that some works have found the zeropoint from \citet{Lindegren2021} to ``overcorrect'' parallaxes. However, this is mainly the case for very bright stars ($G\lesssim 10$). In particular, \citet{Zinn2021} found the correction to perform well for giants with similar colors and magnitudes to the Giraffe, so we expect the corrected {\it Gaia} parallax to be reliable.   

\cite{Jayasinghe2022} noted that the spectrophotometric parallax calculated by \citet{Hogg2019} for the source implies a somewhat larger distance than the parallax. However, the spectral type---luminosity correlation that is leveraged by their spectrophotometric model is unlikely to hold for a stripped giant. Indeed, any spectroscopic distance indicator trained on ``normal'' stars is expected to overestimate the distance to a stripped star: at fixed $T_{\rm eff}$ and $\log g$, luminosity is proportional to mass, so a star that is half as massive as expected will be half as luminous.

\subsubsection{Radius}
\label{sec:radius}
We infer component radii by fitting the SED. We predict bandpass-averaged mean magnitudes for both components using empirically-calibrated theoretical models from the BaSeL library \citep[v2.2;][]{Lejeune1997, Lejeune1998}. We assume a \citet{Cardelli_1989} extinction law with total-to-selective extinction ratio $R_V =3.1$, and we use an extinction prior from the \citet{Green2019} 3D dust map. We use \texttt{pystellibs}\footnote{\href{http://mfouesneau.github.io/docs/pystellibs/}{http://mfouesneau.github.io/docs/pystellibs/}} to interpolate between model spectra, and \texttt{pyphot}\footnote{\href{https://mfouesneau.github.io/docs/pyphot/}{https://mfouesneau.github.io/docs/pyphot/}}  to calculate synthetic photometry. 

We first fit a single-star model, and found an angular diameter $\Theta = 2R/d=65.5\pm 1\,\mu\rm as$ and $T_{\rm eff} = 4200\pm 60\,\rm K$, with extinction $E(B-V)=0.55$. Including the distance uncertainty, this implies $R_{\rm giant} = 26\pm 2\,R_{\odot}$. This is slightly smaller than found by \citet{Jayasinghe2022}, mostly because of the different assumed distance. 

However, a 2nd luminous source contributes in both the optical and APOGEE spectra: our fitting implies a 20\% contribution in the $H$-band, and 30-60\% contributions in the optical. We therefore fit the SED with a sum of two luminous stars, with the additional constraint that the secondary should contribute 20\% of the total light at 16,000\,\AA\. The results of this fitting are shown in Figure~\ref{fig:sed}. The best-fit radius of the giant is slightly smaller -- $(25\pm 1.5)\,R_{\odot}$ -- than when we fit a single-star model. The effective temperature is also lower, and in better agreement with the spectroscopic fit. This fit implies that the secondary contributes an increasing fraction of the total light toward bluer wavelengths, and should dominate the light at $\lambda \lesssim 4,500$\,\AA. This is in good agreement with the flux ratios we infer from the optical spectral (bottom panel of Figure~\ref{fig:sed}; only the $H-$band point was included in the fit). 

\citet{Jayasinghe2022} found that the secondary contributes only $\lesssim 10\%$ of the light in the $r$-band, and contributes negligibly at $\gtrsim 7,000$ \AA. However, this seems unlikely, given that we find a 20\% contribution at 16,000 \AA. This apparent tension is likely primarily a result of how \citet{Jayasinghe2022} measured the contributions from the secondary: rather than fitting the spectra with two components simultaneously, they fit with a single-star model, and then subtracted off the resulting best-fit model. But this model is different from the actual spectrum of the primary, because it tries to fit the light of the secondary as well  \citep[e.g.][]{El-Badry2018_theory}.

\subsection{Light curves and constraints from ellipsoidal variability}
\label{sec:light_curves}

\begin{figure*}
    \centering
    \includegraphics[width=\textwidth]{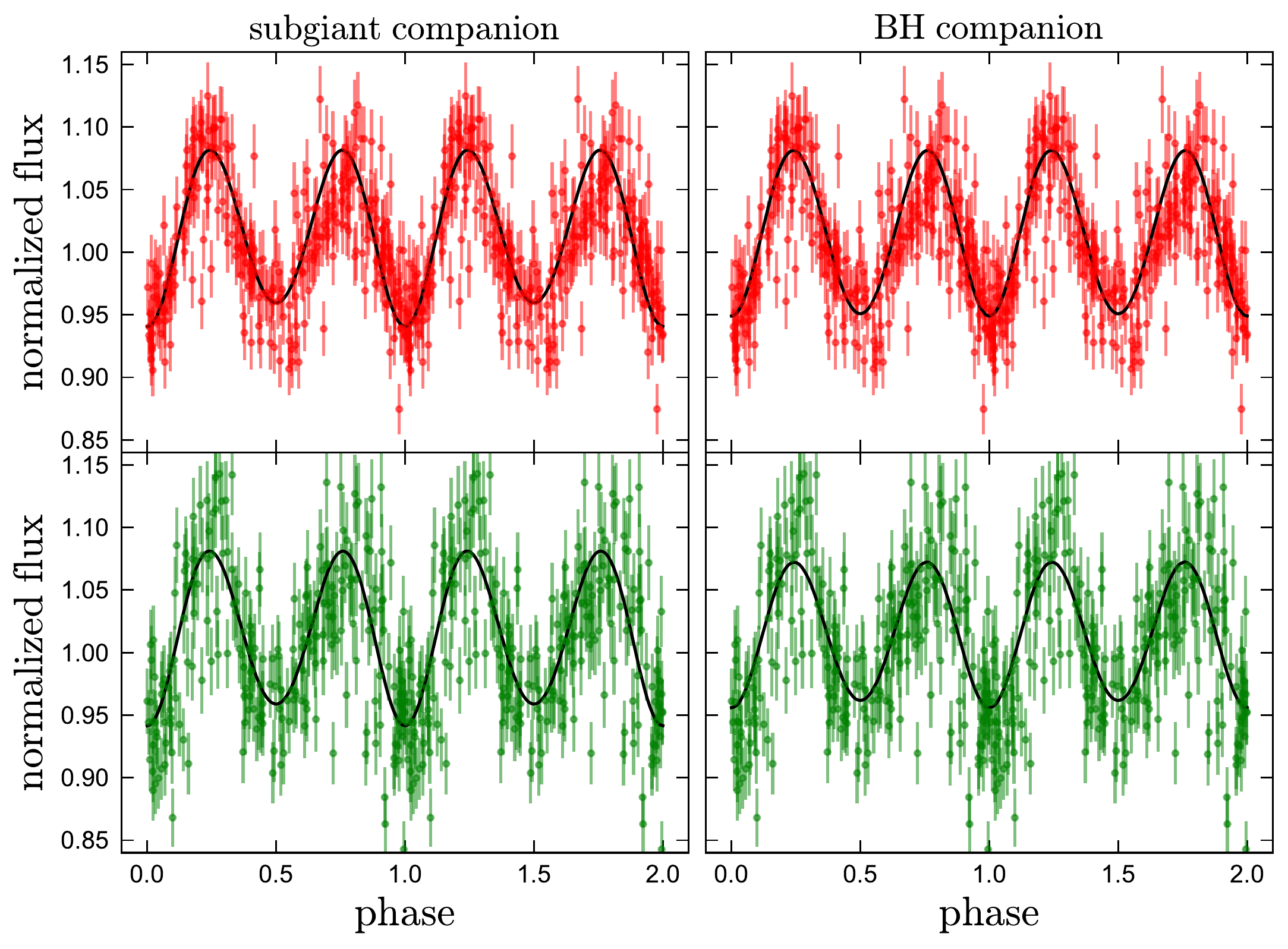}
    \caption{ZTF $r-$band (top) and $g-$band (bottom) light curves of the Giraffe. The observed light curves in the left and right columns are the same. Left column shows PHOEBE prediction for the model we propose: a Roche-lobe filling $\approx 0.35\,M_{\odot}$ giant viewed at $i\approx 50$ deg with dilution from a luminous subgiant companion. Right column shows the model from \citet{Jayasinghe2022}: a not-quite Roche-lobe filling $\approx 0.6\,M_{\odot}$ giant viewed at $i\approx 41$ deg with a dark companion. The model light curves are similar except that gravity darkening (i.e., the asymmetry in adjacent minima) is stronger in the subgiant model, because the inclination and Roche-filling factor are higher. This gravity darkening is also evident in the data, and the subgiant model provides a better fit.  }
    \label{fig:light_curves}
\end{figure*}

We analyzed the Giraffe's $g-$ and $r-$ band light curves from the 8th data release of the Zwicky Transient Facility \citep[ZTF;][]{Bellm2019}; these are shown in Figure~\ref{fig:light_curves}. We focus on ZTF data because it has the smallest empirical photometric uncertainty of the available ground-based light curves. Photometry from {\it TESS} is also available, but it does not span a full orbital period and thus is not well suited for determining the full ellipsoidal variability amplitude and the amount of gravity darkening.

We phased the ZTF data to the RV ephemeris calculated by \citet{Jayasinghe2022} and compared them to a range of models calculated with \texttt{PHOEBE} \citep[][]{prsa2005, prsa2016}. We focus on two main features of the light curves: (a) the total variability amplitude, which provides a joint constraint on inclination, the amount of dilution from the secondary, and the giant's Roche lobe filling factor, and (b) the strength of gravity darkening, as manifest in the different depths of the light curve minima at phase 0 and 0.5.

The min-to-max variability amplitude of the phased light curves is $14\pm 1$\% in both $g-$ and $r-$ bands. One might be tempted to infer that this means there is no more dilution from the secondary in $g$ than in $r$. However, we find that with no dilution, the predicted variability amplitude from \texttt{PHOEBE} is $\approx 20\%$ larger in $g$ than $r$ due to wavelength-dependent limb and gravity darkening. This implies that there is 15-20\% more dilution from the secondary in $g$ than in $r$.

\begin{figure}
    \centering
    \includegraphics[width=\columnwidth]{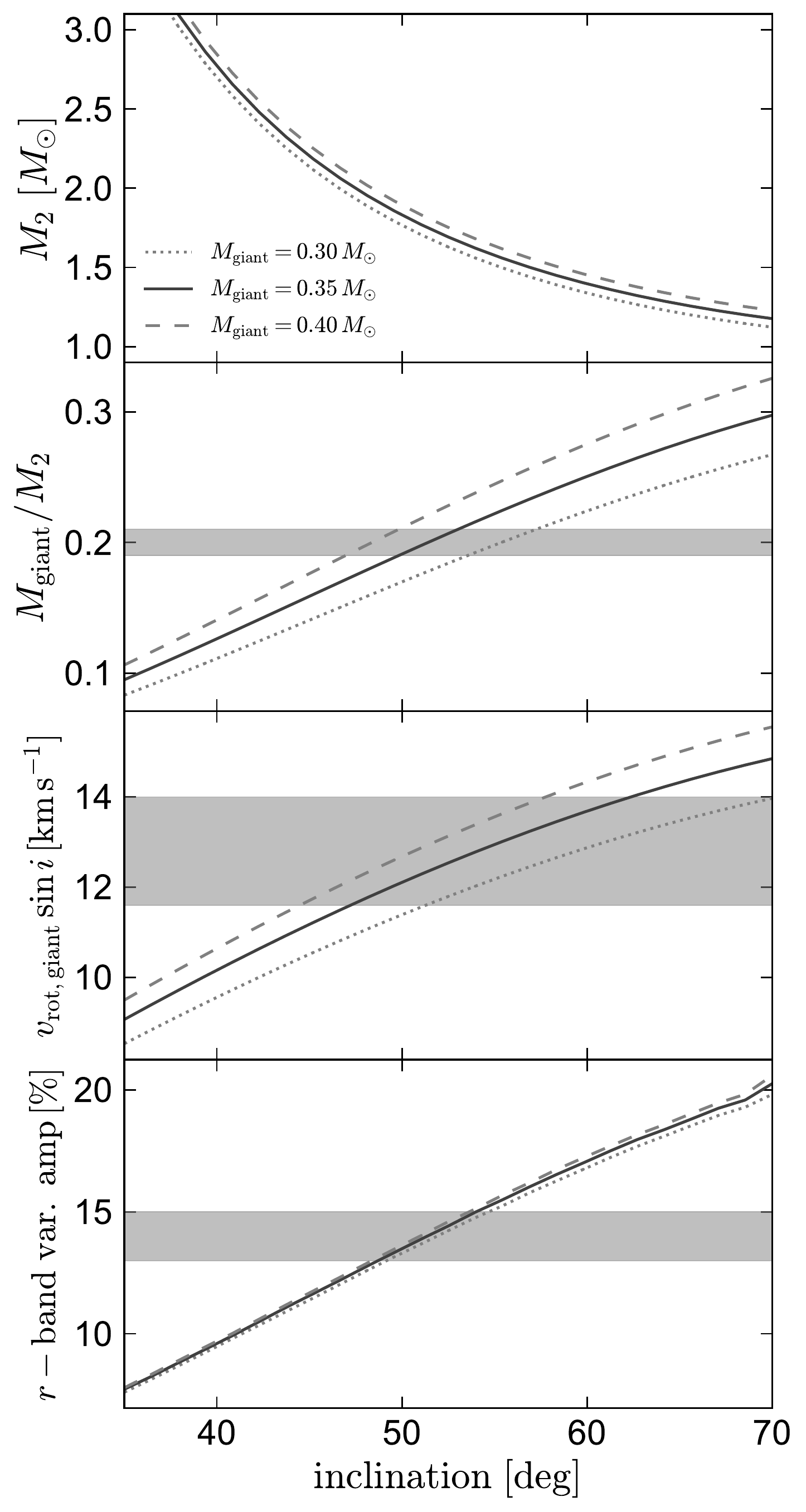}
    \caption{Predicted companion mass, mass ratio, projected rotation velocity, and photometric variability amplitude as a function of inclination, for three different plausible giant masses in the Giraffe. Shaded regions show observational constraints. These are matched when the inclination is between 48 and 56 degrees, corresponding to a companion mass between 1.5 and 2 $M_{\odot}$. The assumption of $0.3 \lesssim M_{\rm giant}/M_{\odot} \lesssim 0.4$ is motivated by the constraint that the giant fills its Roche lobe.}
    \label{fig:dynamics}
\end{figure}

We fix the mass ratio to $q=0.2$ based on the RVs and account for dilution from the secondary by modeling it as a star in \texttt{PHOEBE}, with temperature and radius consistent with the SED and spectral fit. The observed ellipsoidal variability amplitude then limits possible solutions to a one-dimensional slice in the inclination vs. Roche lobe filling factor plane. \footnote{The light curve and mass function alone do not strongly constrain the inclination or mass ratio, because solutions with high inclination and lower filling factor produce nearly identical light curves to solutions with lower inclination and higher filling factor. One can always get a {\it formally} tight constraint with enough data, but systematics (e.g. unaccounted-for spots, and especially contamination from the secondary) will drastically limit the utility of these constraints.}  
If we assume that the giant fills its Roche lobe -- which seems reasonable given that it clearly is close (given the ellipsoidal variability) and there is an accretion disk around the companion seen in H$\alpha$ -- then the observed variability amplitude and RV-inferred mass ratio imply an inclination of $i = 50\pm 3$ degrees (Figure~\ref{fig:dynamics}; this is basically independent of the giant's mass). The uncertainty on this inclination is mainly due to uncertainty in how much dilution there is from the companion, and uncertainty in the observed variability amplitude itself.
For an inclination of 50 degrees, the mass function $f(M_2) = 0.58\,M_{\odot}$ implies a companion mass of $M_2 = 1.83\,M_{\odot}$ if $M_1 = 0.35\,M_{\odot}$ (this value is justified in Section~\ref{sec:obs_summary}). This yields a mass ratio $M_1/M_2 = 0.197$, in  agreement with the RV-inferred mass ratio. 

The model favored by \citet{Jayasinghe2022} has a lower inclination of 41 degrees, and thus would predict a lower ellipsoidal variability amplitude without the contributions of the companion. However, their modeling included significantly less dilution from a luminous secondary (e.g., they inferred a dilution factor of only $7\%$ in the $r-$ band, whereas in our model, it is $\approx 35\%$).  

We proceed under the assumption that the giant fills its Roche lobe.
A scenario in which it does not fill its Roche lobe seems a priori less likely for two reasons. First, it requires more fine tuning, because we would have to observe the giant just after detachment, rather than during the much longer period where there is ongoing mass transfer (Section~\ref{sec:evol}). This is true irrespective of the nature of the companion. Second, the observed double-peaked H$\alpha$ emission suggests that there is a sizable accretion disk, which is not expected in a detached system with only wind mass transfer. %because (a) the accretion rate will be lower (by a factor of $\approx 100$ compared to the mass transfer rated predicted by our models in Section~\ref{sec:evol}), and (b) the angular momentum of the captured wind will be lower.

Figure~\ref{fig:light_curves} compares the ZTF $r-$ and $g-$ band light curves of the Giraffe to PHOEBE model predictions for a luminous subgiant companion (left) and a BH companion (right). The models make similar predictions, because the amplitude of the main signal -- near-sinusoidal variation on half the orbital period due to tidal deformation of the giant -- is covariant between inclination, Roche lobe filling factor, and dilution. There is, however, a key difference between the two models: the predicted asymmetry due to gravity darkening is stronger -- in agreement with the data -- in the case with a luminous companion. This is a result of the higher inclination and Roche lobe filling factor in the subgiant model, and another piece of evidence for the luminous companion. 

Although both light curve models explain most of the observed variability, there are some systematic difference between the shape of the observed light curve and both models. Most obviously, the observed light curve maximum at phase $\approx 0.25$ is higher than the one at phase $\approx 0.75$. The scatter in the phased data is also somewhat larger than expected given the photometric uncertainties, suggesting some long-term variability. Both of these features could be due to spots on the giant, or changes in the disk around the secondary.

\subsection{Summary of the observational constraints}
\label{sec:obs_summary}
Limits on the masses of both components are summarized in Figure~\ref{fig:dynamics}. Given the inferred giant radius of $R_{\rm giant}=(25\pm 1.5)R_{\odot}$, the constraint that it fills its Roche lobe translates to $M_{\rm giant}=(0.33\pm 0.06)\,M_{\odot}$ \citep[e.g.][]{El-Badry2022}. The mass ratio from the companion's reflex motion then implies $M_2$ between $1.38\,M_{\odot}$ and $2\,M_{\odot}$. The inclination required to match the observed mass function over this range ranges for 48 to 58 deg, with higher masses implying lower inclinations. The inclination constraint from the observed ellipsoidal variability amplitude is slightly more stringent (lower panel of Figure~\ref{fig:dynamics}), excluding inclinations above $\approx 56$ deg. We thus conclude (before considering possible evolutionary scenarios) that giant masses ranging from 0.29 to 0.39 are most plausible, corresponding to companion masses ranging from 1.5 to 2 $M_{\odot}$.

\subsubsection{Plausible subgiant secondaries}
\begin{figure}
    \centering
    \includegraphics[width=\columnwidth]{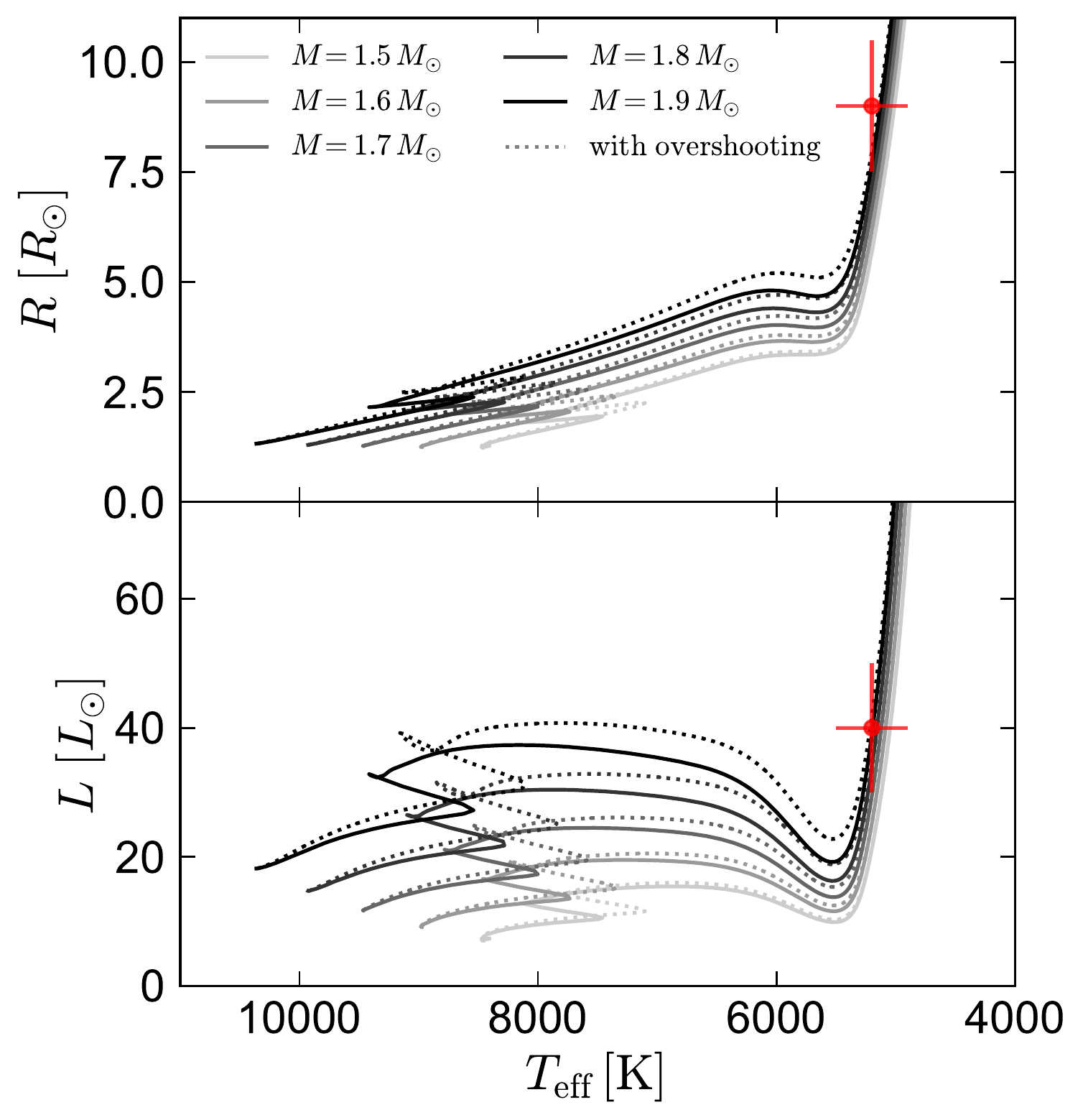}
    \caption{MESA single-star evolutionary tracks with masses compatible with our constraints on the secondary in the Giraffe. Red symbol shows our inferred parameters of the secondary. Dotted and solid lines show models with and without convective overshooting. Models with a relatively wide range of masses ascend the giant branch at similar temperatures. The dynamically-inferred $1.5 \lesssim M_2/M_{\odot} \lesssim 2$ predicts a secondary temperature, radius, and luminosity consistent with the observationally inferred values. }
    \label{fig:subgiants}
\end{figure}

The properties we infer for the companion are consistent with any subgiant in the dynamically-inferred plausible mass range. Figure~\ref{fig:subgiants} compares our constraints on the subgiant's temperature, radius, and luminosity to single-star evolutionary tracks. We calculate these with MESA \citep{Paxton_2011} for two different assumptions above convective overshooting, which can significantly change the models' luminosity in the Hertzsprung gap.\footnote{We use default parameters in MESA version r15140 and initial $Z=0.006$. In the models with overshooting, we use the exponential scheme in all overshooting regions with \texttt{overshoot\_f = 0.014} and \texttt{overshoot\_f0 = 0.004}. Both sets of calculates use \texttt{mixing\_length\_alpha = 2.0} and atmosphere boundary conditions from \citet{Castelli2003}.  } 

Subgiants with a wide range of masses ascend the Hayashi track at similar temperatures -- always with $5000\lesssim T_{\rm eff}/\rm K \lesssim 5500 $. These are always faint in the UV and thus do not conflict with the UV limits. Our observed temperature and luminosity of the secondary thus do not strongly constrain its mass (particularly given that recent accretion could have affected its evolution in a way not captured by single-star models), but there is no tension between the dynamically inferred $M_2 = 1.5-2\,M_{\odot}$ and the properties of the secondary.

\section{The Unicorn}
\label{sec:data_unicorn}

\subsection{Optical spectra}
We analyze the 7 Keck/HIRES spectra of the Unicorn that were also analyzed by \citet{Jayasinghe2021}. These provide useful coverage of most of the range between 3500 and 8000 \AA, with typical per-epoch SNR of 100 per pixel at 5000\,\AA and resolution $R=60,000$. The phase coverage samples most of the dynamic range in the giant's RVs, but does not include observations near $\phi=0.5$ (when the companion would be eclipsed for sufficiently high inclinations).

The qualitative appearance of the spectra changes dramatically from the blue to red orders. Blueward of 3800\,\AA, the spectra have broad absorption lines, indicative of rapid rotation (e.g. Figure~\ref{fig:unicorn_rvs2}). They are well-fit by a single-star model with $T_{\rm eff}\approx 5800\,\rm K$, $\log g \approx 3$, and $v\sin i \approx 70\,\rm km\,s^{-1}$. The source here appears to be almost stationary, with epoch-to-epoch RV variations of $\lesssim 20\,\rm km\,s^{-1}$. On the other hand, the redder regions of the spectra are dominated by narrow lines from a cooler and much more slowly rotating star ($v\sin i\approx 15\,\rm km\,s^{-1}$). This component is much more RV variable, with epoch-to-epoch RV variations exceeding 120\,$\rm km\,s^{-1}$. \citet{Jayasinghe2021} focused their analysis on red wavelengths, so the properties of the giant they report correspond to the object that dominates there.

\subsubsection{Spectral disentangling and fitting}
\label{sec:unicorn_disentangling}

\begin{figure*}
    \centering
    \includegraphics[width=\textwidth]{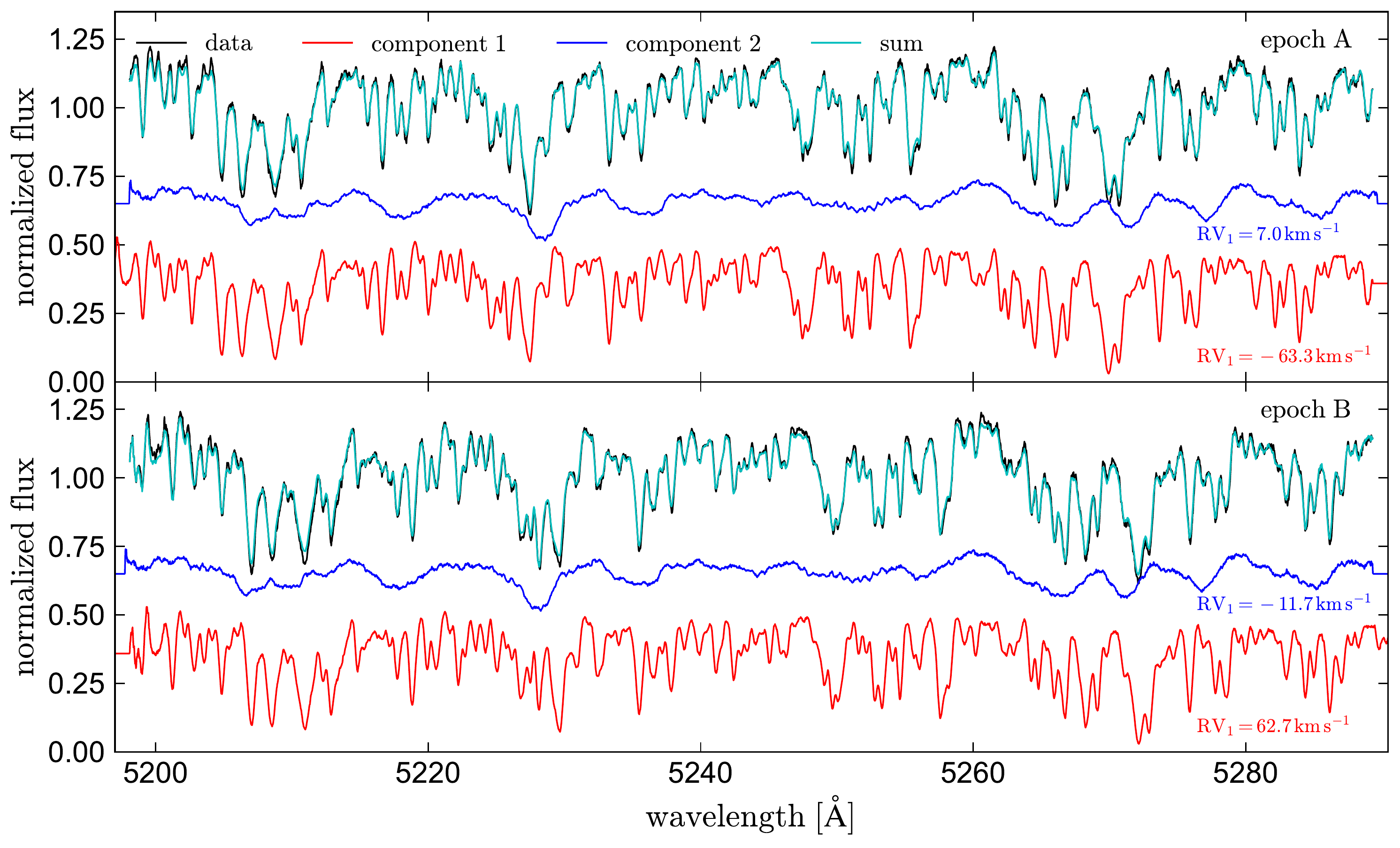}
    \caption{Spectral disentangling of the Unicorn. Top and bottom panels show two different epochs, chosen to be near opposite quadratures. Black lines show the observed Keck/HIRES spectra. Blue and red lines show the disentangled spectra of the giant and subgiant (same in both panels except RV shifts), whose sum (cyan) matches both observed spectra, as well as those in the other 5 epochs that are not shown. The RVs of both components at each epoch are printed in the lower right. The observed spectra are well-modeled as the sum of two components. The subgiant component (blue) contributes $\approx 65\%$ of the light at these wavelengths, but its contributions to the spectra are subtle because its absorption lines are broadened by rapid rotation. }
    \label{fig:unicorn_disentangle}
\end{figure*}

\begin{figure*}
    \centering
    \includegraphics[width=\textwidth]{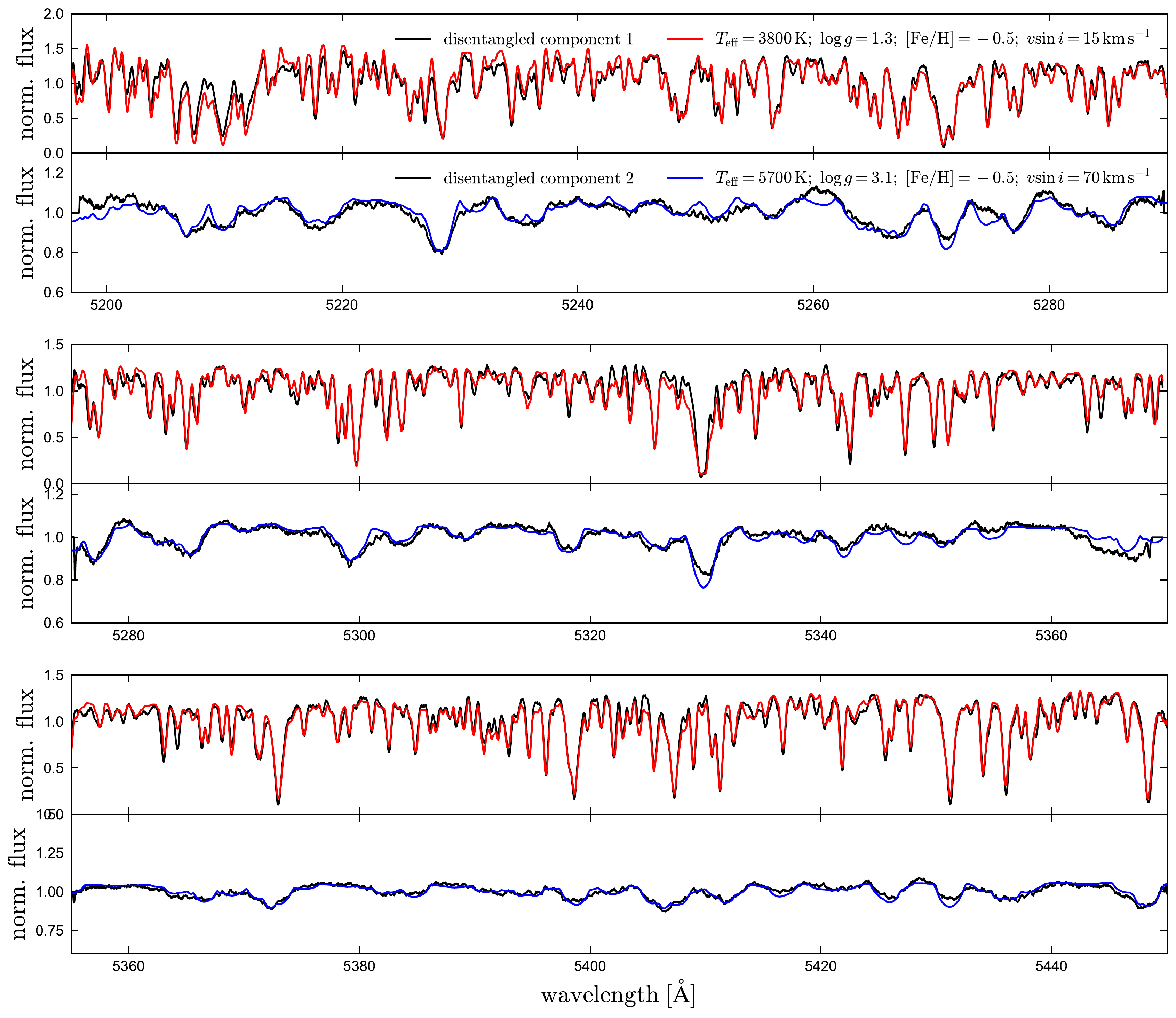}
    \caption{Comparison of disentangled spectra of the Unicorn to models. Black lines show the normalized disentangled spectra of both components; red and blue lines show Kurucz models for a giant and rapidly-rotating subgiant. Three sets of panels show three different spectral regions. The agreement between data and models is generally quite good.  }
    \label{fig:unicorn_models}
\end{figure*}

\begin{figure*}
    \centering
    \includegraphics[width=\textwidth]{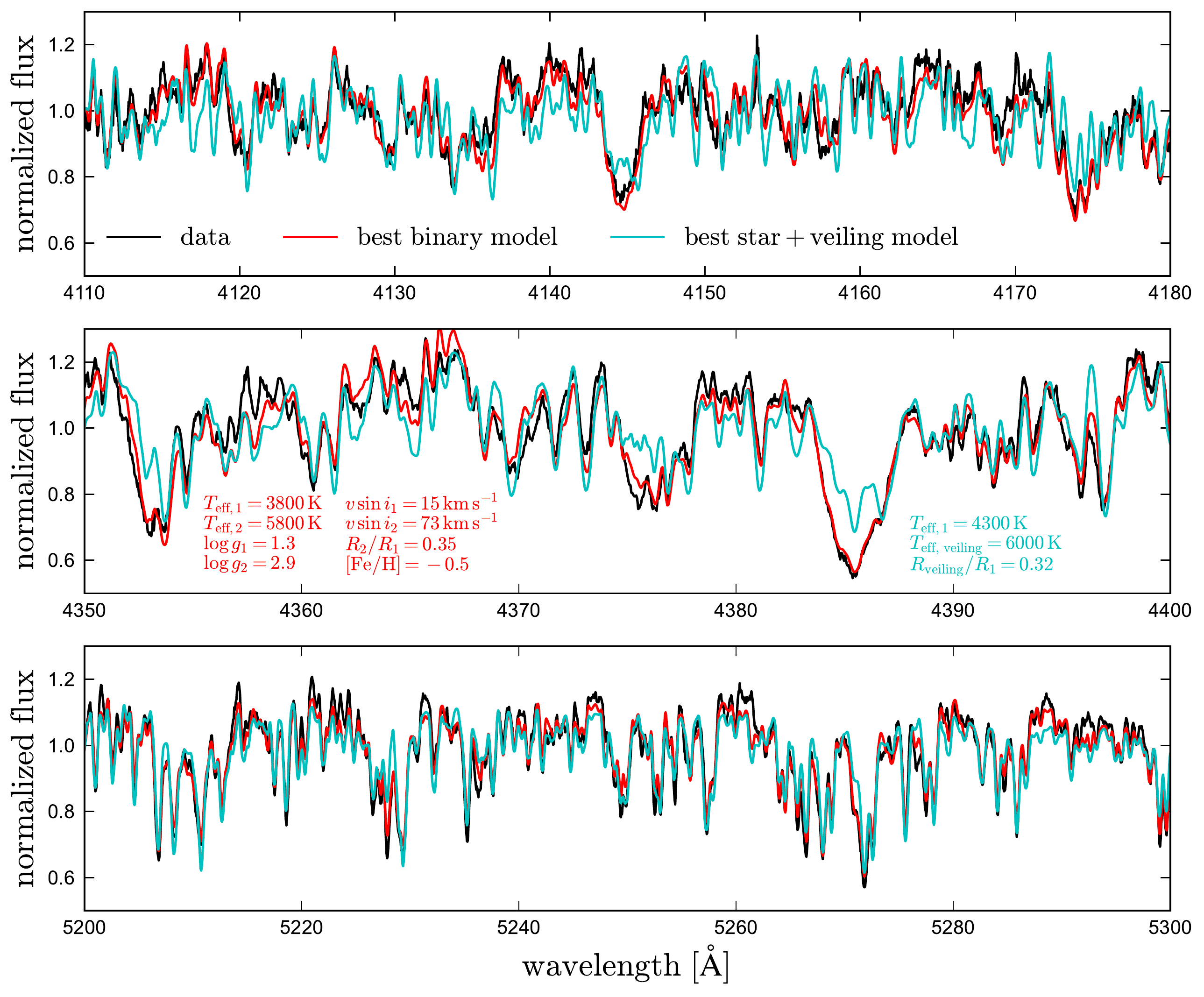}
    \caption{Spectral fits to a single-epoch spectrum of the Unicorn. Black line shows cutouts of the spectrum. Red line shows the best-fit binary model, the parameters of which are listed in the middle panel. Cyan line shows the best-fit single-star + veiling model, in which the secondary is modeled as a blackbody that contributes continuum but no absorption lines. The binary model achieves a far better fit, particularly at blue wavelengths. For the best-fit binary model, the giant (component ``1'') contributes 85\%, 80\%, and 63\% of the light in each panel. Its contributions to the spectrum are less obvious at longer wavelengths due to its higher rotation velocity. The giant dominates the light at $\lambda \gtrsim 6,500$\,\AA.  }
    \label{fig:unicorn}
\end{figure*}
We fit the optical spectra (the Unicorn was not observed by APOGEE) using a binary spectral model. As in the Giraffe, we find that two luminous components contribute to the optical spectra: a giant with effective temperature of $3800 \pm 100$\,K, and a subgiant with  effective temperature $5800 \pm 200$~K. The fractional flux contributions of the subgiant are larger than in the Giraffe. However, its inferred rotation velocity,  $v\sin i \approx 70\,\rm km\,s^{-1}$, is significantly larger, so its spectral lines are not visually obvious in the composite spectra.

We performed spectral disentangling following the same approach used for the Giraffe. Two luminous components are again successful in describing the composite spectra (Figure~\ref{fig:unicorn_disentangle}), and the disentangled spectra are well-fit by synthetic spectral models (Figure~\ref{fig:unicorn_models}). The giant spectrum again shows evidence of CNO processing, with $\rm [C/N]\approx -1.1$.

\citet{Jayasinghe2021} found that although continuum dilution was evident at blue wavelengths, the spectrum of the diluting component appeared smooth, without star-like spectral features. To explore this further, we attempted to fit a single-epoch spectrum of the Unicorn with a single-star model that includes veiling by a second source that contributes only continuum, but no absorption lines. We modeled this source's spectrum as the {\it continuum} from a Kurucz model (without absorption lines, similar to a blackbody). The results are shown in Figure~\ref{fig:unicorn}; the three panels show independently-fit regions of the spectrum. Although a single star + continuum model provides a better fit than a pure single-star model, it performs much worse than a binary model, particularly at blue wavelengths. The shortcomings of the single-star + veiling model become less severe at redder wavelengths, where the secondary contributes less light and has fewer strong lines. Across all orders and epochs, the $\chi^2$ difference between the best-fit binary and single star + veiling models is more than $10^6$.

As with the Giraffe, we find flux contributions of the secondary that are larger than estimated previously.  At 7,000\,\AA, where \citet{Jayasinghe2021} found the veiling to become negligible, the secondary still contributes 45\% of the light. Indeed, the total luminosity of the secondary, $L_2 \approx 70\,L_{\odot}$ is comparable to the luminosity of the giant, $L_1 \approx 105\,L_{\odot}$. The best-fit temperature we find for the giant, $T_{\rm eff,\,giant}\approx 3,800\,\rm K$, is significantly cooler than the value of 4,400\,K found by \citet{Jayasinghe2021}.

\subsubsection{Variations in the H-alpha line}
\label{sec:halpha_unicorn}
\citet{Jayasinghe2021} found that the H$\alpha$ absorption line in the giant changes in width and depth as a function of orbital phase, and becomes deepest at $\phi \approx 0.5$, when the companion would be behind the giant. They interpreted this as the result of emission from an accretion disk around the companion, which in their model would be eclipsed at $\phi=0.5$. We investigated the origin of this variability using the Keck/HIRES spectra. 

We initially found that the observed H$\alpha$ absorption line of the giant is significantly deeper than predicted by the best-fit Kurucz model. We suspected, however, that this was mainly a result of the fact that ab-initio spectral models generically fail to reliably predict the Balmer lines and other strong lines in cool giants \citep[e.g.][their Figure 2]{Bergemann2016}. To test this, we tried modeling the spectrum of the giant with an appropriately broadened empirical spectrum of a giant with $T_{\rm eff}\approx 3800\,\rm K$, $\log g \approx 1.2$, and $[\rm Fe/H]=-0.5$ from the GALAH survey \citep[][]{Buder2021}. When combined with a model spectrum for the subgiant, this yielded a good fit to all 7 composite spectra. The composite line profile varies significantly from epoch to epoch, but this is mainly because there are two sets of spectral lines moving back and forth. We thus found no strong evidence of emission in the Balmer lines during the ``out-of-eclipse'' observations. 

However, \citet{Jayasinghe2021} did find that the H$\alpha$ line at $\phi \approx 0.5$ becomes significantly deeper than at other phases (their Figure 14). One possible explanation is that this is due to an accretion stream from the giant's inner Lagrange point to the secondary, which would move between the subgiant and Earth at phase 0.5. Variations in the secondary's spectrum with phase will be explored further in future work (Jayasinghe et al., in prep).

\subsubsection{Dynamical mass ratio}
\label{sec:dynmassratio}
\begin{figure*}
    \centering
    \includegraphics[width=\textwidth]{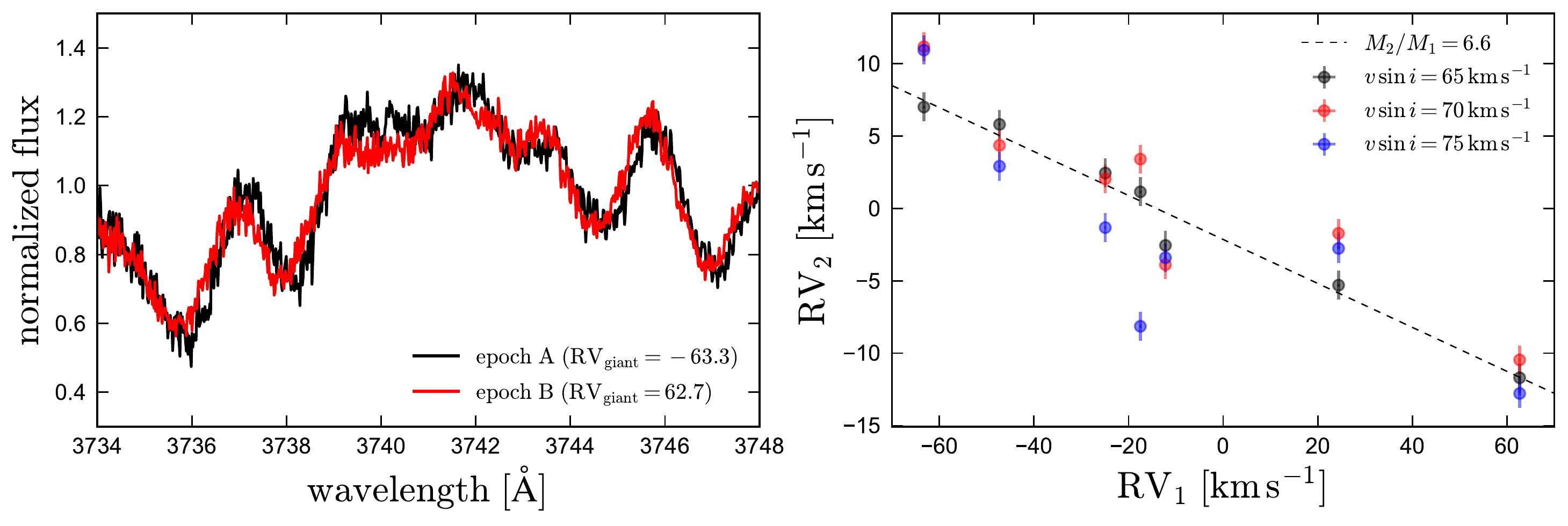}
    \caption{Left: Comparison of two single-epoch spectra of the Unicorn at relatively blue wavelengths, where the subgiant dominates. Although the lines are broad, there is definite evidence of RV variability. Right: RVs for the secondary inferred by cross-correlating different subgiant model spectra with the observed blue spectra. The inferred RVs are somewhat sensitive to the adopted model spectra (and to the spectral region considered), but strongly suggest that the two components  move in anti-phase, with $M_{\rm giant}/M_2 \lesssim 0.17$. }
    \label{fig:unicorn_rvs2}
\end{figure*}

We also explored constraints on the Unicorn's mass ratio from the reflex motion of the companion. In contrast to the slowly-rotating subgiant in the Giraffe, the companion in the Unicorn has broad lines, making measurement of its RVs nontrivial. We attempted several different methods.

First, we cross-correlated a subgiant template spectrum against the bluest few HIRES orders, where the contributions of the giant are negligible. This yielded RVs that unambiguously move in anti-phase with the giant (Figure~\ref{fig:unicorn_rvs2}). However, the implied RV semi-amplitude is sensitive to the choice of subgiant template and $v\,\sin i$, with different plausible templates yielding semi-amplitudes ranging from 4 to 11\,$\rm km\,s^{-1}$. Using an observed single-epoch spectrum instead of a synthetic template resulted in a semi-amplitude of $\approx 6$\,$\rm km\,s^{-1}$, but the precise value varied with the wavelength range being fit. Finally, we fit for the subgiant semi-amplitude during disentangling of the redder orders, where both components contribute comparable fractions of the light. This yielded values ranging from 0 to 11 $\rm km\,s^{-1}$, again with significant variation between orders.

Inspecting the bluest orders, we find that there are non-negligible changes in the companion's line profiles from epoch to epoch. We suspect that these variations, perhaps caused by mass transfer and/or changes in the giant's light-weighted effective temperature due to gravity darkening, are partly at fault for the poorly determined mass ratio. They likely also contribute to the unusual RV curve for the secondary inferred by \citet{Strassmeier2012}, who interpreted it as evidence that the system is a triple. We conclude that further work is required to pin down the mass ratio, though our analysis suggest it is highly unequal, with $K_{2}\lesssim 11\,\rm km\,s^{-1}$ and thus $M_{\rm giant}/M_{2}\lesssim 0.17$. A compact triple still seems unlikely, given the small RV variations of the secondary.

\subsection{Spectral energy distribution and radius}

\begin{figure*}
    \centering
    \includegraphics[width=\textwidth]{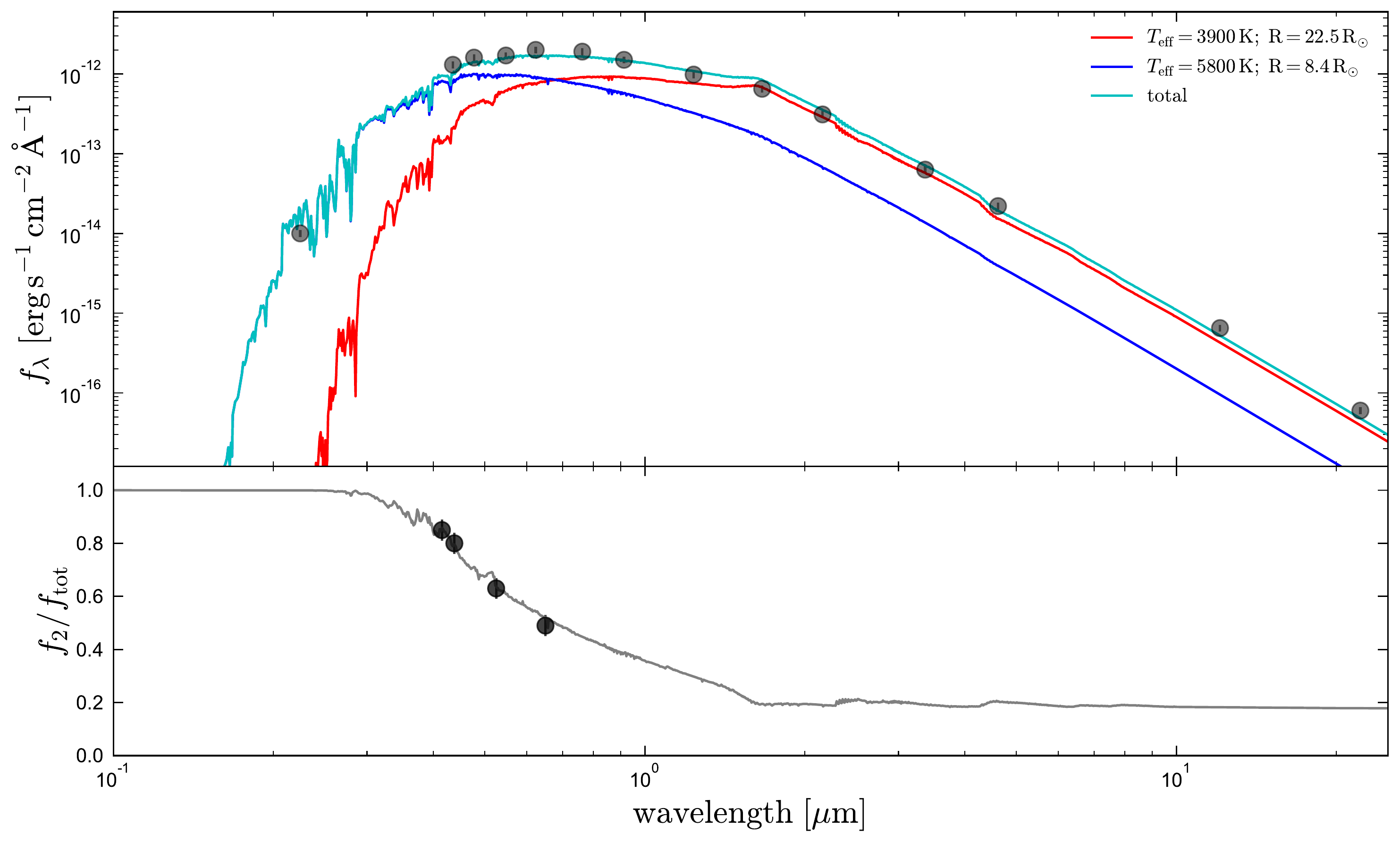}
    \caption{Spectral energy distribution of the Unicorn. Gray points show observed fluxes; red and blue lines show model spectra for the giant and subgiant. We assume a distance of 454 pc and reddening $E(B-V)=0.15$. Bottom panel shows the fraction of the total light that is contributed by the subgiant; gray points show measurements of this quantity from spectral fitting. }
    \label{fig:unicorn_sed}
\end{figure*}

We infer the radii of both components from the SED, following the same approach used for the Giraffe. We do not include the SkyMapper photometry, which is likely saturated. The results are shown in Figure~\ref{fig:unicorn_sed}. The inferred giant radius is $R\approx 22.5\pm 1\,R_{\odot}$, and the effective temperatures of both components are consistent with our spectroscopic fits. The subgiant is predicted to contribute $\approx 20\%$ of the light at $\lambda > 16,000$\,\AA, also similar to the Giraffe.

\subsection{Ellipsoidal variability and mass constraints}
The large-amplitude ellipsoidal variability observed in the Unicorn implies that the giant must nearly fill its Roche lobe, once dilution from the companion is accounted for. For example, the observed min-to-max variability amplitude in the $V-$band is $\approx 17\%$, but a (presumably) barely-variable companion contributes $\approx 60\%$ of the $V-$band light. This implies that the min-to-max variability amplitude of the giant is $\approx 37\%$, which implies a Roche lobe filling factor $R_{\rm giant}/R_{\rm Roche\,lobe}\gtrsim 0.97$. As in the Giraffe, we interpret this and the time-variability of the Balmer lines as evidence for ongoing mass transfer, and proceed under the assumption that the giant fills its Roche lobe.

In this case, the orbital period and observed giant radius constrain the giant mass to $M_{\rm giant} = 0.44\pm 0.06\,M_{\odot}$. The inclination is constrained by the amplitude of ellipsoidal variability and the lack of eclipses. For plausible giant and subgiant parameters, we find that inclinations $i\lesssim 62\,\rm$ deg are ruled out because they would underpredict the observed ellipsoidal variability amplitude even with a Roche lobe-filling giant. Inclinations $i > 72\,\rm deg$ would produce detectable eclipses and are similarly ruled out. 

\citet{Jayasinghe2021} interpreted the increased Balmer absorption at $\phi \approx 0.5$ (Section~\ref{sec:halpha_unicorn}) as evidence for eclipses of the companion and a near edge-on inclination. Such a scenario is difficult to reconcile with the secondary detected in the spectra. Irrespective of the nature of the secondary, the spectra imply that it is an optically thick source with $T_{\rm eff}\approx 5800\,\rm K$ and surface area similar to an $R \approx 8\,R_{\odot}$ sphere. Such a companion would cause eclipses in the light curve for edge-on inclinations, even if it were e.g. a disk around a BH. Given our constraints on the secondary $T_{\rm eff}$ and optical luminosity, it also cannot be a larger circumbinary structure: this would have much larger luminosity and would outshine the giant at all wavelengths.

We adopt a fiducial inclination of $i=70$ deg, which matches the observed variability amplitude for our fiducial parameters, but we consider values from 62 to 72 deg plausible. For $i=70$ deg, our PHOEBE models predict ellipsoidal variability amplitudes of 9.5\% in the $B$ band and $17.2\%$ in the {\it TESS T}  band; both of these are in good agreement with the observed values.

Given the observed mass function, $f(M_2)=1.71\,M_{\odot}$, these masses and inclinations correspond to companion masses $2.60\lesssim M_2/M_{\odot} \lesssim 3.3$, with $M_2\approx 2.8\,M_{\odot}$ for our fiducial model. This is not far from the range inferred by \citet{Jayasinghe2021}, because our lower assumed $M_{\rm giant}$ and lower assumed inclination move the implied companion mass in opposite directions. 

\subsubsection{Tidal RV}
Independent of ellipsoidal variability, another diagnostic of the giant's tidal distortion is the ``tidal RV''; i.e., the apparent deviation from a purely Keplerian orbit due to distortion of the giant's surface. This was investigated for the Unicorn by \citet{Masuda2021}. 

In the regime where $M_2 \gg M_{\rm giant}$, the tidal RV at fixed orbital period scales as $K_{\rm tidal}\sim 1/\rho_{\rm giant}\sin^2(i)$, where $\rho_{\rm giant}$ is the giant's mean density. Compared to the \citet{Jayasinghe2021} model, the model we propose has lower inclination ($62-72$ deg vs 87 deg) and lower $\rho_{\rm giant}$ by a factor of ~2. These factors work in opposite directions, but the $\rho_{\rm giant}$ factor wins out, such that before contributions of the secondary are taken into account, our model implies a $K_{\rm tidal}$ $\approx 50\%$ larger than their model. This is in somewhat worse apparent agreement with the data than the \citet{Jayasinghe2021} model, though the \citet{Masuda2021} analysis also implies a low giant mass ($\approx 0.6\,M_{\odot}$) when the smaller giant radius we find is taken into account. 

The giant's inferred RVs may, however, be biased due to the presence of a 2nd peak to the cross-correlation function (CCF). To investigate this, we generated mock composite spectra matching our inferred parameters for the Unicorn and computed CCFs using the giant's spectrum as a template. The rapidly-rotating subgiant adds a broad second peak to the CCF and pulls the CCF-inferred giant RVs toward the subgiant, leading to a maximum bias of $\approx 1.5\,\rm km\,s^{-1}$. This bias is large compared to the tidal RV amplitude and threatens to overwhelm the tidal signal. However, the bias varies approximately sinusoidally with phase, and so much of it is absorbed into the giant's inferred orbit (meaning that $K_{\rm giant}$ is underestimated by $\approx 1\,\rm km\,s^{-1}$). Once this is accounted for, the maximum bias in the tidal RV due to an unaccounted-for secondary is about $0.5\,\rm km\,s^{-1}$, with the exact value depending on the region of the spectrum considered and the parameters of both stars. We conclude that contributions of a luminous secondary lead to non-negligible changes in the inferred tidal RV and most likely explain the modest tension between our model and the tidal distortion measured by \citet{Masuda2021}.

\subsubsection{Comparison to single-star models}
\label{sec:unicorn_single_star}

\begin{figure}
    \centering
    \includegraphics[width=\columnwidth]{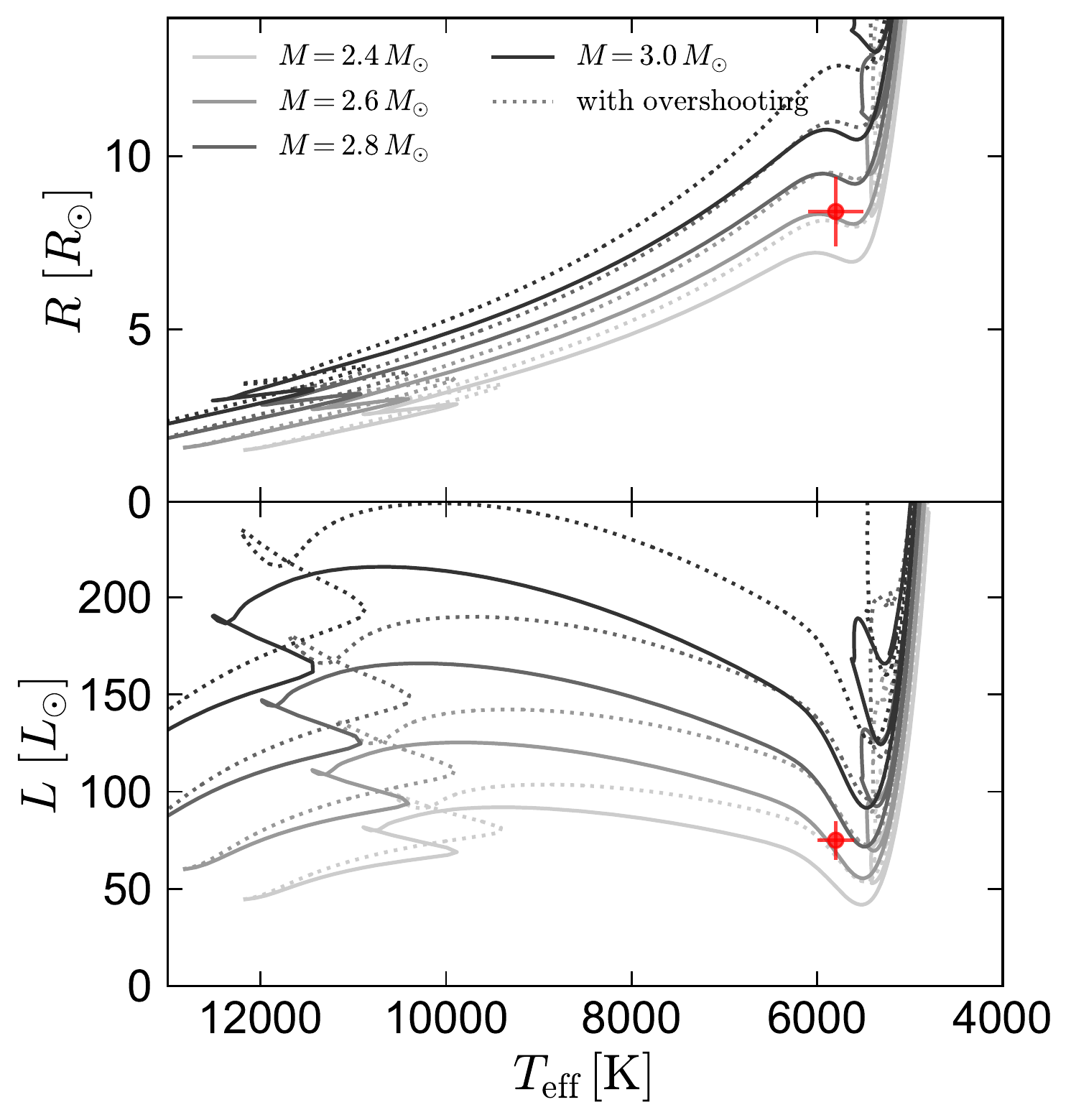}
    \caption{MESA single-star evolutionary tracks with masses compatible with our constraints on the secondary in the Unicorn. Red symbol shows our inferred parameters of the secondary. Dotted and solid lines show models with and without convective overshooting. The dynamically-implied mass of the secondary is $M_2 \approx 2.8\pm 0.2 M_{\odot}$. This is consistent with the tracks without overshooting. The tracks with overshooting imply a somewhat lower mass,   $M_2 \approx 2.4\pm 0.2 M_{\odot}$. }
    \label{fig:subgiants_unicorn}
\end{figure}

In Figure~\ref{fig:subgiants_unicorn}, we compare our measured temperature, radius, and luminosity for the secondary to MESA single-star evolutionary models for subgiants with a range of masses. We again show tracks with and without convective overshooting. the Unicorn has a secondary in a region of the HR diagram where choices related to mixing during the main-sequence evolution (especially overshooting and the adopted mixing length $\alpha$) significantly affect models' evolution \citep[e.g.][their Figures 14 and 15]{Choi_2016}. The secondary mass implied by the tracks without overshooting is $\approx 2.7\,M_{\odot}$, which is consistent with our measurement. On the other hand, the tracks with overshooting imply a significantly lower mass, $\approx 2.4\,M_{\odot}$. The reason for this is that overshooting increases the mass of the convective core during main-sequence evolution, and thus the mass of the core at hydrogen exhaustion. A larger core mass leads to a larger radius and luminosity.

The assumption of “no overshooting” is somewhat extreme, and so we conclude that there is modest tension between the dynamical and single-star evolutionary mass of the secondary, which may be a result of binary interactions. However, the differences between the two sets of models shown in Figure~\ref{fig:subgiants_unicorn} reflects the fact that that detailed evolutionary tracks are uncertain at the companion's current evolutionary stage, and it is not implausible that uncertainties in the stellar evolutionary calculations (e.g., mixing length, rapid rotation, detailed abundances of the accreted envelope) can account for the apparent tension.

\section{Evolutionary state: low-mass giants on the way to becoming helium white dwarfs}
\label{sec:evol}

Low-mass helium white dwarfs (He WDs) are the cores of giants stripped through binary mass transfer before central helium ignition. Hundreds of such WDs have been discovered in the last decade as companions to main-sequence stars, other WDs, and neutron stars \citep[e.g.][]{Brown2010, Maxted2014, Masuda2019, El-Badry2021lamostj, El-Badry2021_preelm}. Most helium WDs in wide orbits ($P_{\rm orb}\gtrsim$ few hours) are thought to form via stable mass transfer, with the giant's envelope being gradually stripped over tens of Myr as it ascends the giant branch and the orbit expands. During the mass transfer process, the donor stars are low-mass giants, with masses of order $0.2-0.5\,M_{\odot}$. A few such systems have been observed during the stripping process \citep[e.g.][]{Miller2021}. Once almost all of the giant's envelope has been removed, mass transfer ceases and what is left of the giant contracts, ``freezing in'' the current orbital period.

We propose that the giants in the Giraffe and Unicorn are currently in this stripping phase. This evolutionary scenario is qualitatively similar to the one proposed to explain the BH candidates LB-1, HR 6819, and NGC 1850 BH1 \citep[e.g.][]{Shenar2020, El-Badry2021hr6819, El-Badry2022}. The main difference is that the donors in the Giraffe and Unicorn (a) are at an earlier evolutionary stage and (b) have an even lower mass than in those systems, and thus will most likely evolve directly to helium WDs rather than igniting core helium burning and becoming sdOB stars first. This however depends sensitively on the masses of the giants \citep[e.g.][]{Heber2016}, and a future for them as low-mass sdOB stars is also not ruled out.

A prototypical example of a system with the proposed evolutionary history is the binary Regulus, which at $V=1.4$ is one of the $\approx 20$ brightest stars in the sky. That system contains a proto-helium WD in a 40-day orbit with a main-sequence companion \citep{Gies2008, Gies2020}. Both component masses in Regulus are consistent, within their uncertainties, with our measurements for the Unicorn. The system's evolutionary history is likely very similar to that of the Unicorn and Giraffe \citep[][]{Rappaport2009}, except that mass transfer has already ended and the stripped giant's core has contracted.

Because there is a tight correlation between the radius of a giant (which sets the orbital separation and period) and the mass of its degenerate core, one expects a tight correlation between the mass of helium WDs formed by stable mass transfer and orbital period \citep[e.g.][]{Rappaport1995, Tauris1999, Lin2011, Chen2017}. This relation -- which predicts a core mass of $\approx 0.3\,M_{\odot}$ for detached systems at periods of 60-80 days -- has been mostly validated by observations \citep{Tauris2014}.  For systems in which mass transfer is still ongoing, the giant mass can be higher, so $M \gtrsim 0.3\,M_{\odot}$ is a lower limit on the expected giant mass in this scenario.

\subsection{Binary evolution models}
We calculated a small grid of binary evolution models using MESA \citep[Modules for Experiments in Stellar Astrophysics, version \texttt{r12778};][]{Paxton_2011, Paxton_2013, Paxton_2015, Paxton_2018, Paxton_2019}.  MESA simultaneously solves the 1D stellar structure equations for both stars, while accounting for mass and angular momentum transfer using simplified prescriptions. 

We assumed an initial composition of $X=0.712, Y=0.28, Z=0.006$ for both stars. Opacities are taken from OPAL  \citep{Iglesias1996} at $\log(T/{\rm K}) \geq 3.8$ and from \citet{Ferguson2005} at $\log(T/{\rm K}) < 3.8$. Nuclear reaction rates are from \citet{Angulo1999} and \citet{Caughlan1988}. We use the \texttt{photosphere} atmosphere table, which uses atmosphere models from \citet{Hauschildt1999} and \citet{Castelli2003}. The \texttt{MESAbinary} module is described in \citet{Paxton_2015}. We used the \texttt{evolve\_both\_stars} inlists in the MESA test suite as a starting point for our calculations, and most inlist parameters are set to their default values. Roche lobe radii are computed using the fit of \citet{Eggleton_1983}. Mass transfer rates in Roche lobe overflowing systems are determined following the prescription of \citet{Kolb_1990}. The orbital separation evolves such that the binary's total angular momentum is conserved, as described in \citet{Paxton_2015}. 

We first highlight models that approximately reproduce the observed parameters of the Giraffe. A wide range of initial binary parameters can produce a $0.3-0.4\,M_{\odot}$ Roche-lobe filling giant with an orbital period of $\sim$80 days \citep[e.g.][]{Kalomeni2016}. Here we explore only a subset of the possible parameter space. We initialize the calculation with a detached binary containing two zero-age-main sequence stars on a circular orbit. We experimented with initial periods ranging from 7 to 20 days, initial primary masses ranging from 1 to 2 $M_{\odot}$, and initial mass ratios ranging from 0.8 to 1. We model mass transfer as being fully conservative for simplicity. This is a reasonable assumption during most of the calculation, but it is unlikely to hold during the initial period of thermal-timescale mass transfer, when the mass transfer rate briefly exceeds the thermal limit by a factor of $\sim$10. 

We present one model in detail in Figure~\ref{fig:mesa_params} and~\ref{fig:mesa}, and comment on the effects of varying initial parameters in Section~\ref{sec:var_params}. The model we highlight has initial primary mass $1.15\,M_{\odot}$, initial mass ratio 0.997, and initial period of 13 days. (We comment on the assumption of initial mass ratio $q\approx 1$ in Section~\ref{sec:fine_tune}.) The evolution of the primary, which is today the giant, is shown in Figure~\ref{fig:mesa_params}. Mass transfer begins as the primary is on its way up the giant branch, when its radius is $11\,R_{\odot}$. Because of the nearly-equal mass ratio, the secondary has also already left the main-sequence at that time, with a radius of $6.5\,R_{\odot}$.

Mass transfer is at first rapid, proceeding on the donor's thermal timescale and reaching $\dot{M}\approx 3\times 10^{-4}\,M_{\odot}\,\rm yr^{-1}$, but slows once the giant has become the less-massive component, such that the mass transfer leads to orbit widening. Subsequent evolution is governed by the donor's nuclear evolution, and the system gradually expands as a semi-detached binary for $\approx$50\,Myr. It reaches an orbital period of 81 days $\approx 45$ Myr after the onset of mass transfer, when the giant's mass and radius are $M_{\rm giant}=0.37$ and $R_{\rm giant} =24.3\,R_{\odot}$. At this time the secondary mass and radius are $M_2 = 1.93\,M_{\odot}$ and $R_2 = 10\,R_{\odot}$, yielding a mass ratio $M_{\rm giant}/M_2 = 0.19$. The predicted surface carbon-to-nitrogen ratio is $\rm [C/N] =-1.0$, in good agreement with the observed value. This is a consequence of material that was inside the giant's convective core during its main-sequence evolution becoming exposed at the surface once the envelope is stripped: without mass transfer, the surface [C/N] ratio for a $\approx 1\,M_{\odot}$ giant would be near 0 \citep[e.g.][]{Hasselquist2019}.

A qualitatively similar scenario can produce the Unicorn, but the total initial mass must be larger in order to produce the observed final mass. For near-conservative mass transfer, the current system masses imply an initial primary mass between about 1.6 and 1.9 $M_{\odot}$; values up to $\approx 3\,M_{\odot}$ are possible if mass transfer is highly nonconservative. We refer to \citet{Rappaport2009} for a detailed exploration of the initial masses and periods required to produce such a system, which are quite similar to those that can produce the Regulus system.

\subsubsection{Future evolution}

\begin{figure}
    \centering
    \includegraphics[width=\columnwidth]{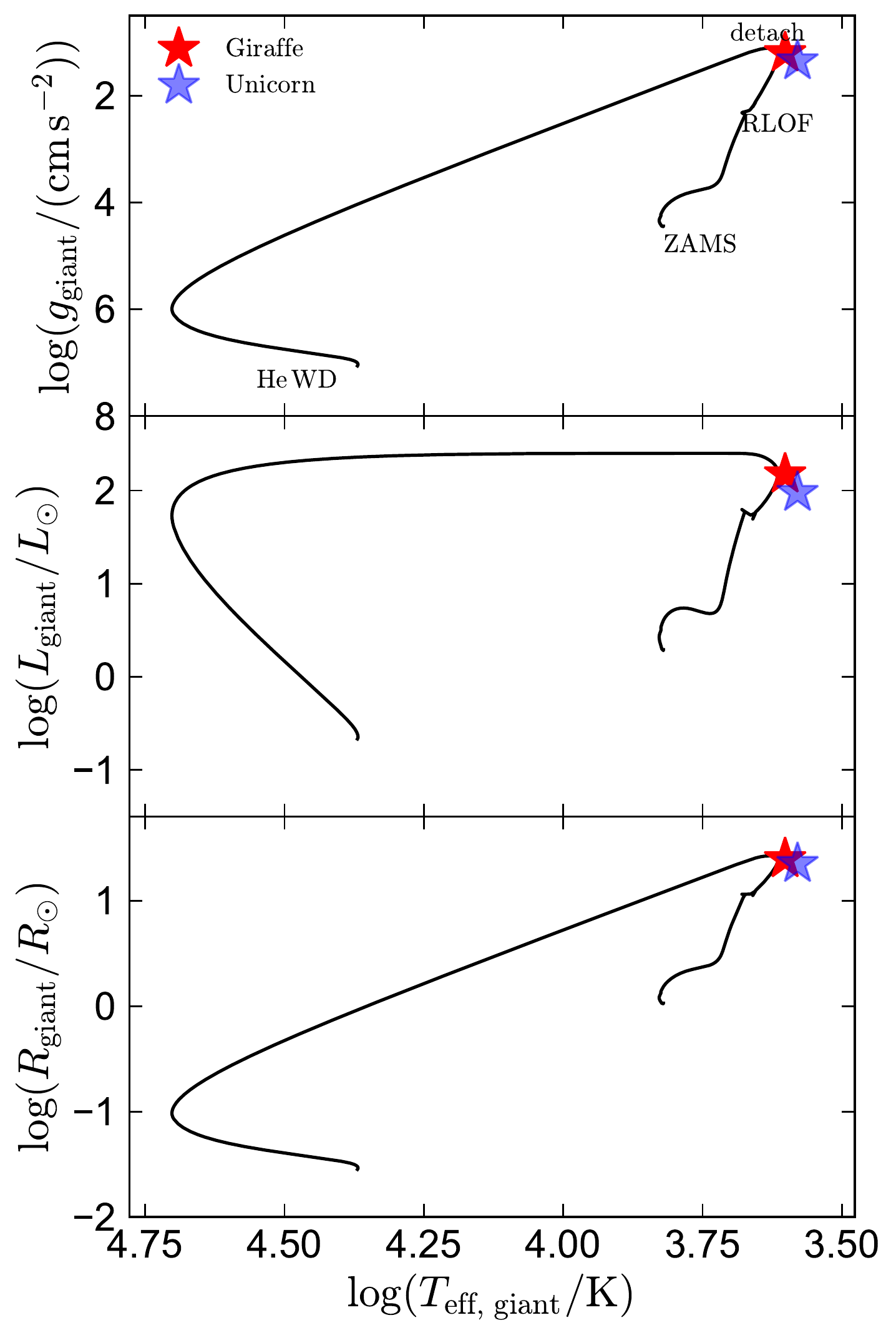}
    \caption{Evolution of the giant in a MESA model similar to the Giraffe. Red and blue stars shows the observed giants in the Giraffe and Unicorn, with uncertainties comparable to the size of the symbols. Evolution begins at the zero-age main sequence (ZAMS). Roche lobe overflow (RLOF) occurs at a period of 13 days, as the primary is ascending the giant branch. The donor continues to lose mass via stable RLOF as it expands, and it matches the properties of the Giraffe near the tip of the giant branch. Once almost all of the envelope has been lost to the companion, the donor rapidly contracts and heats up, terminating its evolution as a low-mass helium white dwarf (He WD) with an orbital period of $\approx 100$ days.}
    \label{fig:mesa_params}
\end{figure}

\begin{figure*}
    \centering
    \includegraphics[width=\textwidth]{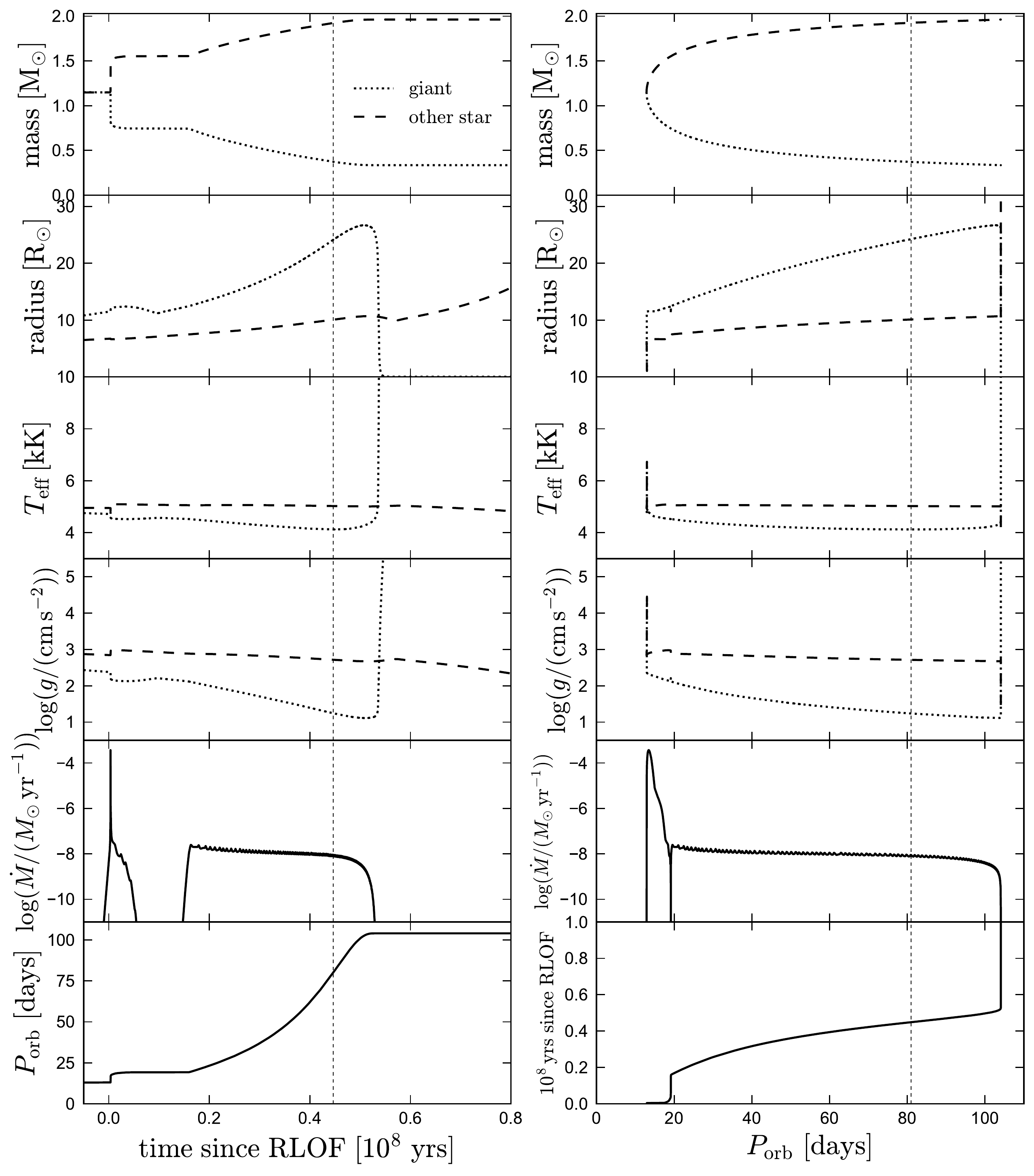}
    \caption{Properties of the MESA model from Figure~\ref{fig:mesa_params} as a function of time since first Roche lobe overflow (left) and orbital period (right). Dashed vertical line marks a period of 81 days, when the model's parameters are similar to those observed in the Giraffe. The lifetime of the current phase, with a low-mass giant undergoing stable Roche lobe overflow to a subgiant companion, is $\approx 30$\,Myr.  }
    \label{fig:mesa}
\end{figure*}

The orbit of the model shown in Figure~\ref{fig:mesa_params} continues to expand until it reaches a period of 104 days. At this point, the giant is a $\approx 0.32\,M_{\odot}$ helium core, with a $\approx 0.006\,M_{\odot}$ diffuse hydrogen envelope. Mass transfer ceases and the giant rapidly contracts and heats up, leaving behind a $\approx 0.33\,M_{\odot}$ helium WD. Although they do not occur in this calculation, we note that one or more shell flashes may occur as the proto-He WD reaches the cooling track \citep[e.g.][]{Driebe1998, Istrate2014}.

We terminate the calculation once the He WD settles in its cooling track. Eventually, the companion will ascend the giant branch until it overflows its Roche lobe at a radius of order $30\,R_{\odot}$. This will likely lead to an episode of common envelope evolution, most likely leaving behind a compact binary containing two He WDs or a sdOB + He WD binary \citep[e.g.][]{Tutukov1988, Nelemans2000}. Such binaries have recently been discovered in significant numbers via light curve variability \citep[e.g.][]{Burdge2020}, and many more are expected to be found by {\it LISA}. 

Depending on the orbital period after common envelope ejection, a sdOB + He WD binary may begin another episode of mass transfer while the sdOB star is still burning helium, evolving into an AM CVn phase \citep[e.g][]{Brooks2015}. If the binary comes in contact when both components are already degenerate, the most likely outcome is a merger leaving behind a single sdOB star \citep{Schwab2018}. The details are uncertain because they depend sensitively on the component masses and post-common-envelope separation (and the outcome of compact WD binary interactions is uncertain even for known masses; e.g., \citealt{Shen2015}), but in any case, both systems likely have an exciting future.

\subsection{Varying model parameters}
\label{sec:var_params}
Here we consider what range of initial parameters could produce systems like the Giraffe and Unicorn. 

\begin{figure}
    \centering
    \includegraphics[width=\columnwidth]{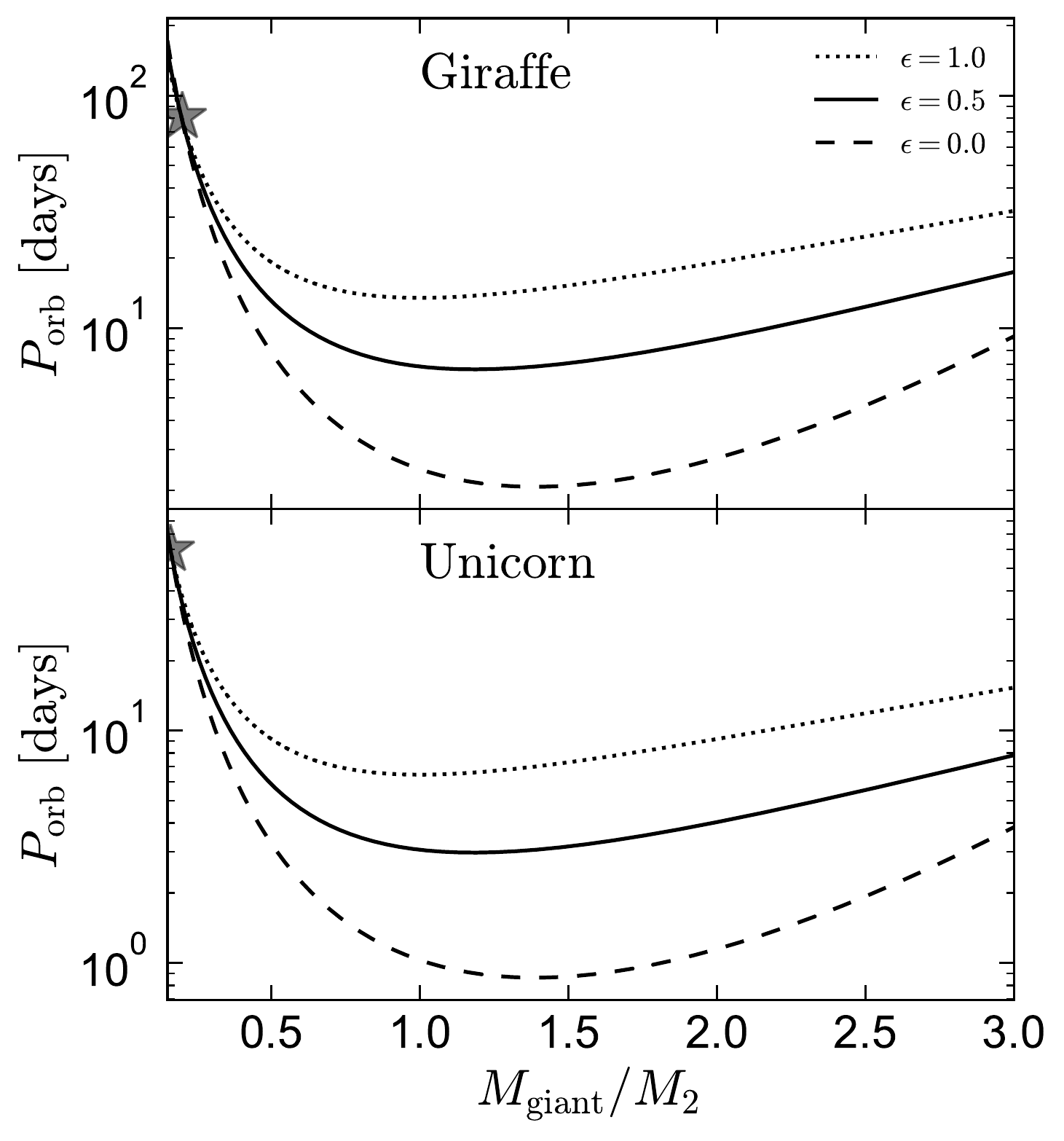}
    \caption{Orbital evolution of the Giraffe (top) and Unicorn (bottom). Dotted line shows fully-conservative mass transfer. Dashed lines shows fully non-conservative mass transfer, with the assumption that the ejected mass carries away the specific angular momentum of the accretor. Solid line shows a 50\% accretion efficiency with the same assumption. Gray stars show the observed periods and mass ratios. Assuming an initial mass ratio near 1, the Giraffe's initial period would be $\approx 13$ days for conservative mass transfer and $\approx 2.5$ days for fully non-conservative. For the Unicorn, the same values are $\approx 6$ days and $\approx 1$ day. }
    \label{fig:P_inits}
\end{figure}

\begin{itemize}
    \item {\it Initial primary mass}: A relatively wide range of initial primary masses can produce a stripped, Roche-lobe filling $0.3-0.4\,M_{\odot}$ giant in a 60- or 81-day orbit. This is because the radii of giants with degenerate cores is set mainly by core mass \citep[e.g.][]{Rappaport1995}. Primaries with larger initial masses begin mass transfer earlier in their evolution up the giant branch, but their masses and radii subsequently follow similar tracks to lower-mass stars at a given orbital period.
    
    The total mass of the Giraffe today is constrained to $2\lesssim M_{\rm tot}/M_{\rm odot} \lesssim 2.5$, and that of the Unicorn is constrained to $3.1\lesssim M_{\rm tot}/M_{\rm odot} \lesssim 3.8$. This limits the initial primary mass to $1 \lesssim M_{1}/M_{\rm odot} \lesssim 1.3$ (Giraffe) and $1.6 \lesssim M_{1}/M_{\rm odot} \lesssim 1.9$ (Unicorn), unless mass transfer is significantly non-conservative. Highly nonconservative mass transfer is not implausible: many of the models we explore go through an initial period of very rapid thermal-timescale mass transfer, with $\dot{M}$ up to $10^{-3}\,M_{\odot}\,\rm yr^{-1}$ when a subgiant donor first overflows its Roche lobe.

    \item {\it Initial mass ratio}: If the mass ratio at the onset of Roche lobe overflow is too unequal, the giant will be vulnerable to runaway dynamical mass transfer and a common envelope event. The critical mass ratio beyond which this occurs depends on the structure of the star and remains imperfectly understood \citep[e.g.][]{Hjellming1987, Soberman1997, Pavlovskii2015}. We found mass transfer in our calculations to be stable for initial mass ratios $M_{\rm giant}/M_{2} \lesssim 1.15$ at the onset of RLOF, with some dependence on initial period. However, only initial mass ratios very near unity allow the companion to evolve off the main sequence prior to the onset of mass transfer (Section~\ref{sec:fine_tune}).
    
    \item {\it Initial orbital period}: Longer initial periods cause mass transfer to begin when the donor is more evolved and has had time to form a larger helium core. This leads to a somewhat larger giant mass and radius when the giant reaches a period of 60 or 81 days. If the efficiency of mass transfer is modeled as constant, there are simple analytic relations between the current and initial periods and mass ratios \citep[e.g.][their Equation 29]{Soberman1997}. These allow us to model the systems' period evolution backward in time given constraints on their current orbital periods and mass ratios. The results are shown in Figure~\ref{fig:P_inits} for both systems. For the Giraffe, we assume $M_{\rm giant}/M_2 = 0.198$, as is observationally well-constrained. For the Unicorn, we assume $M_{\rm giant}/M_2 = 0.16$; this is consistent with our constraints but less certain (Section~\ref{sec:dynmassratio}). We expect that the initial mass ratios of both systems were near 1 in order for mass transfer to remain stable and the companion to have time to leave the main sequence. For conservative mass transfer, this implies initial periods of 13 and 6 days in the Giraffe and Unicorn. Non-conservative mass transfer leads to shorter initial periods, down to a few days.
\end{itemize}

\subsection{Rotation rate of the subgiants}
\label{sec:subgiant_rotation}

An interesting difference between the Unicorn and the Giraffe is that the companion in the Unicorn is rotating rapidly (for a subgiant), almost certainly a consequence of recent accretion, while the companion to the Giraffe is rotating slowly. Rapid rotation is a natural prediction of the evolutionary scenario we propose, and so the Giraffe's slow rotation seems more puzzling than the Unicorn's rapid rotation.  There are, however, other cases of mass transfer binaries with highly unequal mass ratios (implying the donor has lost a large fraction of its mass) and slowly-rotating accretors \citep[e.g.][]{Whelan2021}.

Even in the Giraffe, the subgiant is rotating too rapidly to be tidally synchronized. Tides may, however, play a roll in slowing both subgiants' rotation. Both are well inside their Roche lobes, with $R/R_{\rm Roche\,lobe}\approx 0.17$. In the theory of \citet{Zahn1977}, the tidal synchronization timescale due to turbulent friction in the stars' convective envelopes is $t_{{\rm sync}}\approx\beta/\left(6q^{2}k_{2}\right)\left(MR^{2}/L\right)^{1/3}\left(a/R\right)^{6}$. Here $q=M_{\rm giant}/M$, $\beta = I/(MR^2)$, where $I$, $M$, $R$, and $L$ are the moment of inertia, mass, radius, and luminosity of the subgiant, $k_2$ is it apsidal motion constant, and $a$ is the binary semi-major axis. From MESA models of subgiants, we find $\beta \approx 0.025$ in the Unicorn and $\beta \approx 0.14$ in the Giraffe. From \citet{Claret2019}, we estimate $k_2 \approx 0.0015$ for the Unicorn and $k_2\approx 0.07$ for the Giraffe. This leads to $t_{\rm sync} \approx 10\,\rm Myr$ for the Giraffe (shorter than the duration of mass transfer; Figure~\ref{fig:mesa}), and $t_{\rm sync} \approx 100\,\rm Myr$ for the Unicorn (longer). 

It thus seems plausible that tides have played a significant roll in spinning-down the Giraffe, but do not yet play a significant roll in the Unicorn. The difference in the timescales estimated for the two systems is, however, driven mainly by differences in $k_2$ and $\beta$, which are calculated purely from evolutionary models.

\subsection{Comparison to other similar systems}
In Figure~\ref{fig:graveyard}, we compare the Giraffe and Unicorn to several binaries in related evolutionary stages. HR 6819 \citep{Bodensteiner2020, El-Badry2021hr6819} and LB-1 \citep{Shenar2020} are recently-detached systems in which the stripped star is still bloated. NGC-1850 BH1 \citep{El-Badry2022} and HD 15124 \citep{El-Badry2022hd15124} are still mass-transferring. HZ 22 \citep{Greenstein1973} and 59 Cyg \citep{Peters2013} contain stripped stars that have already contracted and heated significantly. These objects are chosen as a representative, but not exhaustive, subsample of the stripped binary population; see e.g. \citet{Schootemeijer2018} and  \citet{Wang2021} for other similar systems.

Figure~\ref{fig:graveyard} also shows the MESA model from Figures~\ref{fig:mesa_params} and~\ref{fig:mesa}, as well as a model for a more massive donor with initial mass of $4\,M_{\odot}$. This model is described in \citet{El-Badry2021hr6819} and approximately reproduces the observed properties of the HR 6819 system. The qualitative evolution of the two models is similar. The most important difference is that a more massive donor ignites helium burning in its core while mass transfer is still ongoing and subsequently becomes a (non-degenerate) sdOB star rather than a He WD. Even higher-mass donors are expected to follow similar trajectories at higher luminosity \citep[e.g.][]{Gotberg2018}. The Unicorn and Giraffe have the coolest and most diffuse donors of the highly-stripped objects shown here. Irrespective of whether their future evolution involves core helium ignition, the models shown in Figure~\ref{fig:graveyard} imply that they will soon contract and heat up.
%Because their masses are near the minimum value for core helium ignition ($\approx 0.33\,M_{\odot}$; e.g., \citealt{Heber2016}), it is unclear whether they will 

\section{Summary and discussion}
\label{sec:discussion}

\begin{table*}
\caption{Summary of constraints.}
\label{tab:summary}
\begin{tabular}{llllllllllll}
system            & $T_{\rm eff,\,giant}$ & $R_{\rm giant}$ & $v\sin i_{{\rm giant}}$ & $M_{\rm giant}$ & $T_{\rm eff,2}$ & $R_2$ & $v\sin i_{{2}}$         & $i$     & $M_2$ & [Fe/H] & [C/N]  \\
\hline
                     & [kK]      &  [$R_{\odot}$]  & $[\rm km\,s^{-1}]$   & $[M_{\odot}]$ & [kK] & [$R_{\odot}$] & $[\rm km\,s^{-1}]$ & [deg]  & [$M_{\odot}$]  & [dex]      & [dex]    \\
\hline
Giraffe     & $4.00 \pm 0.1 $   & $25\pm 1.5 $ & $13 \pm 1 $ & $0.33\pm 0.06$  & $5.15\pm 0.2$ & $9.0\pm 0.5$  & $16\pm 3$ &  $52\pm 4$ & $1.8\pm 0.2$ & $-0.55 \pm 0.1$  & $-1.1 \pm 0.2$  \\
Unicorn & $3.80 \pm 0.1 $ & $22.5 \pm 1.0 $  & $15 \pm 2$ & $0.44\pm 0.06$ & $5.80\pm 0.2$  & $8.3\pm 0.4$ & $70\pm 10$  &  $68\pm4$  & $2.8\pm0.3$       &     $-0.51 \pm 0.1$ & $-1.1 \pm 0.2$  \\
\end{tabular}
\end{table*}

\begin{figure}
    \centering
    \includegraphics[width=\columnwidth]{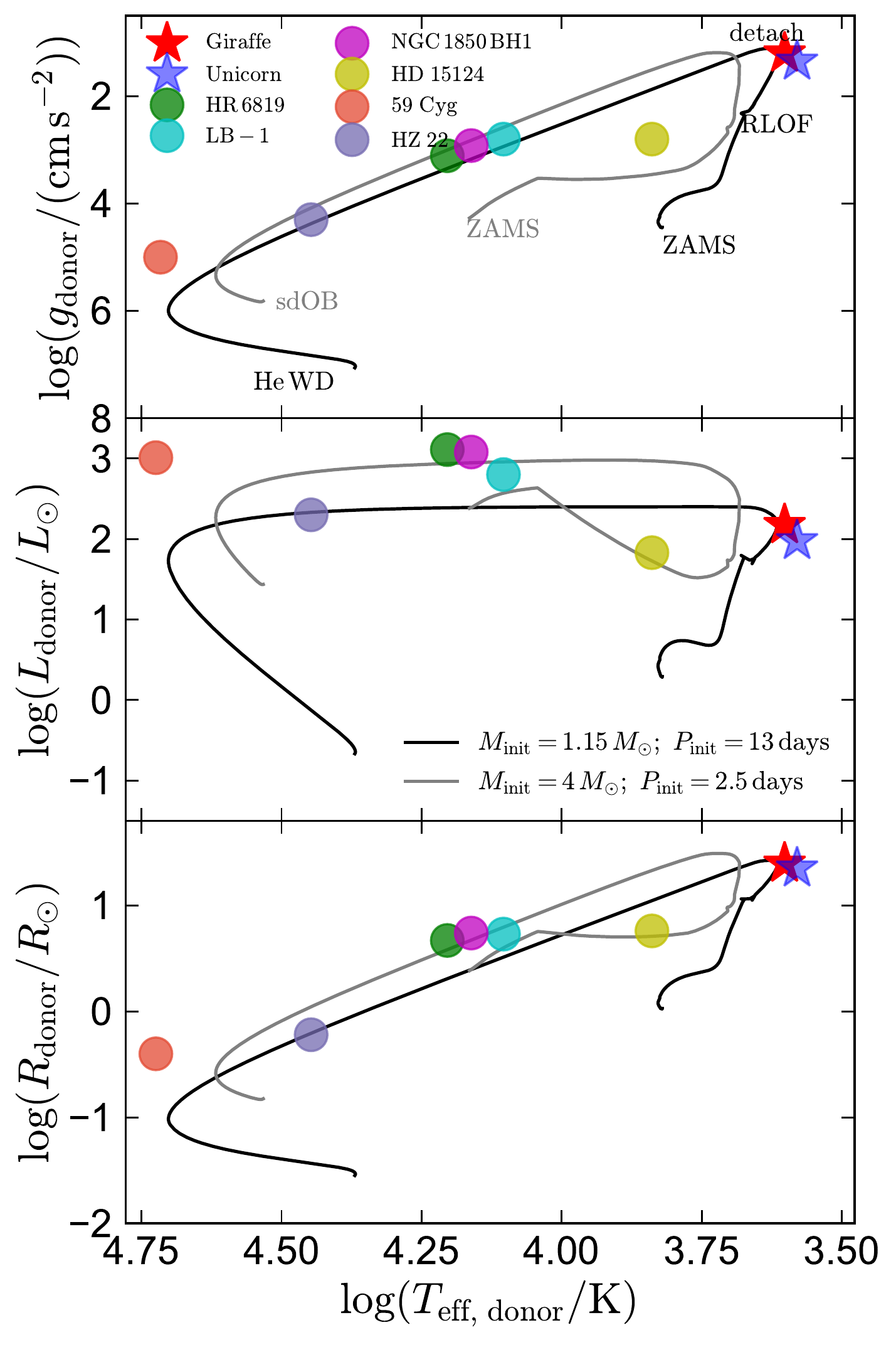}
    \caption{Comparison of the  Giraffe and Unicorn donors (star symbols) to a menagerie of binaries with similar evolutionary histories. Black line shows a MESA model with a relatively low-mass donor that will directly become a He WD (same as Figure~\ref{fig:mesa_params}); gray line shows a higher-mass model that will first become a core helium burning sdOB star. 59 Cyg and HZ 22 are objects that have already begun to contract toward sdOB and He WD fates, respectively. The other objects have donors that are still inflated. These MESA models do not match the detailed properties of all the observed objects, given the diversity of observed periods and masses, but they are representative of the evolutionary histories that produce low-mass, stripped stellar cores via stable mass transfer.}
    \label{fig:graveyard}
\end{figure}

We have re-assessed the BH candidates  2M04123153+6738486 (``the Giraffe'') and V723 Mon (``the Unicorn''), which both contain red giant stars in wide orbits with companions of unknown nature. We find that both systems' properties are best explained by a stripped low-mass ($0.3-0.5\,M_{\odot}$) giant with a luminous subgiant companion. Our constraints for both systems are summarized in Table~\ref{tab:summary}. Like several other recent BH candidates that likely contain two luminous stars \citep[e.g.][]{Shenar2020, Bodensteiner2020, El-Badry2021hr6819, El-Badry2022}, these systems contain stripped stars with masses significantly lower than would be inferred from single-star evolutionary models.

The most important difference between our analysis and previous work is that we fit both systems with binary spectral models. In contrast, \citet{Jayasinghe2021, Jayasinghe2022} fit a single-star model and then analyzed the residuals. This approach leads to biases in the inferred parameters of both components, because the best-fitting single-star spectral model is different from the spectrum of either component. Our main results are as follows:

\begin{itemize}
\item {\it Luminous companions}: Using spectral disentangling, we show that a second luminous source contributes to the optical spectra of both the Giraffe (Figure~\ref{fig:giraffe_disentangle}) and the Unicorn (Figure~\ref{fig:unicorn_disentangle}). In the Giraffe, the companion is also detected in the near-infrared (Figure~\ref{fig:apogee_summary}). In both cases, the companion spectrum is well-fit by a stellar atmosphere model of a subgiant (Figures~\ref{fig:giraffe_models} and~\ref{fig:unicorn_models}).  Our fitting implies that the flux contribution of the secondary is significantly larger than assumed previously: the secondaries in both systems contribute about half the light in the optical, and they dominate at the bluest observed wavelengths (Figures~\ref{fig:sed} and~\ref{fig:unicorn_sed}).

\item {\it Inferred system parameters}:
Dilution from the secondary reduces the amplitude of ellipsoidal variability (Figure~\ref{fig:light_curves}), such that the giants in both systems must be nearly Roche lobe filling. Given the signatures of mass transfer in the spectra, we assume both systems are semi-detached, with ongoing stable mass transfer from the giants to the companions. 

Accounting for contributions from a luminous companion, we find that the observed light curves and SED imply a giant mass of $0.3-0.4\,M_{\odot}$ and a companion mass of $1.5-2.0\,M_{\odot}$ in the Giraffe (Figure~\ref{fig:dynamics}). Given our inferred effective temperature and radius of the secondary, such a companion mass is consistent with single-star models for subgiants (Figure~\ref{fig:subgiants}).

In the Unicorn, we also find a low-mass giant, with $M_{\rm giant} =0.44 \pm 0.06\,M_{\odot}$. The companion is more massive than in the Giraffe, with $M_{2} = 2.8 \pm 0.2\,M_{\odot}$. This dynamical mass may be larger by 10-20\% than the value predicted by single-star evolutionary models given the companion's radius and temperature, potentially a consequence of recent rapid accretion. However, this depends sensitively on the adopted stellar models (Figure~\ref{fig:subgiants_unicorn}).

Both systems have sub-solar metallicity, $\rm [Fe/H]\approx -0.5$, and are strongly enhanced in nitrogen at the expense of carbon compared to the solar abundance pattern, with $[\rm C/N]\approx -1.1$. We interpret this as evidence of CNO processing in the giants' cores prior to envelope stripping. 

\item {\it Evolutionary history}: We can reproduce the observed properties of both systems with binary evolution calculations in which the giant is a low-mass stripped object undergoing stable mass transfer to a companion on its way to becoming a helium white dwarf or low-mass core helium-burning sdOB star in a wide orbit (Figures~\ref{fig:mesa_params} and~\ref{fig:mesa}). This makes both systems likely progenitors to Regulus-type binaries containing a helium WD in a wide orbit around a normal star; these are longer-period versions of El CVn binaries \citep[][]{Maxted2014, vanRoestel2018}. This evolutionary history is quite similar to that proposed for several other recently deposed BH candidates (Figure~\ref{fig:graveyard}).  Eventually, the subgiant companions will evolve and overflow their Roche lobes, most likely leaving behind a compact binary containing two He WDs, a WD and an sdOB star, or two sdOB stars.

\end{itemize}

\subsection{Could the secondaries be associated with accretion onto a BH?}

There are several reasons to believe that the secondaries we detect in optical and IR spectra are not accretion disks around BHs, as was proposed for the Giraffe by \citet{Jayasinghe2022}. There are abundant observations of astrophysical disks in similar contexts to the one proposed for the Giraffe -- in low-mass X-ray binaries, cataclysmic variables, Be stars, and T Tauri stars -- and their spectra generally do not look like slowly-rotating stellar photospheres. They are usually dominated by emission lines, not absorption lines \citep[e.g.][]{Shahbaz1996, Warner_2003}. If absorption lines are present in a disk around a compact object, they should be broadened by at least the Keplerian velocity at the radius where most of the light is emitted. The observed $v \sin i \lesssim 20\,\rm km\,s^{-1}$ for the secondary in the Giraffe -- if it traced material in orbit around a $\approx 3\,M_{\odot}$ BH -- would correspond to material on AU scales -- significantly larger than the binary separation. Such material could exist in a circumbinary disk, but it would then not be expected to trace the secondary's center of mass, as it is observed to do.

An accretion disk that emits significantly in the optical but not in X-rays is also unexpected. Irrespective of whether accretion is radiatively efficient or not, we expect $L_X \gtrsim L_{\rm opt}$ for accretion disks around compact objects at the relevant accretion rates \citep[e.g.][]{Quataert1999}. For example, the LMXB studied by \citet{Strader2016} -- a rare system with an absorption spectrum from the disk at some epochs -- has $L_X \approx 100 L_{\rm opt}$. For the Unicorn and Giraffe, the X-ray upper limits imply $L_X \lesssim 10^{-5} L_{\rm opt}$ and $L_X \lesssim 10^{-3} L_{\rm opt}$. The implied radiative efficiencies in both systems are also much lower than expected for realistic models of accretion flows \citep[e.g.][]{Sharma2007}.

\subsection{Subgiant companions: fine-tuning required?}

\label{sec:fine_tune}

The basic evolutionary scenario we propose for the giants in the Giraffe and Unicorn can occur for a relatively broad range of initial binary masses and separations, and the existence of helium WDs as wide companions to main-sequence stars implies that it is not uncommon in nature. There is one aspect, however, which had to be fine-tuned in our MESA models: in order for the secondary to be observed as a subgiant today, the initial mass ratio has to be very close to 1 (initial $q \gtrsim 0.995$ in the Giraffe, and  $q \gtrsim 0.98$ in the Unicorn). Otherwise, the two stars would not terminate their main-sequence evolution at the same time, and the secondary would be observed as a main-sequence star rather than a subgiant. Unless the secondary has already formed a helium core and started to expand when mass transfer begins, accretion will rejuvenate it.

On the other hand, there is a strong selection effect against systems with non-subgiant accretors: they will have blue/UV excess, making the companions straightforward to detect. That is, if the companions in the Giraffe and Unicorn were still on the main-sequence, these systems would not have been identified as BH candidates in the first place. Nearly-equal mass binaries are not uncommon \citep[e.g.][]{Tokovinin2000, El-Badry2019twin} and are required to produce any observed binary containing two evolved stars. At least one similar system with a subgiant accretor has been observed \citep{Miller2021}.

One possibility that could solve the fine-tuning ``problem" (if indeed it exists) is that the accreting star may not actually be near the end of its main-sequence evolution, but could be only temporarily inflated due to recent accretion on a timescale short enough that its envelope did not have time to thermally adjust. Such temporary expansion does indeed occur in several of the MESA models we explored with main-sequence accretors and high initial mass transfer rates. In our models, these accretors reach equilibrium again before the giant is stripped down to $\approx 0.4\,M_{\odot}$, but it is worth exploring whether there is a plausible model in which the recent mass transfer rate was higher and the accretors are only temporarily inflated. We defer investigation of such a scenario to future work.

%Our basic hermeneutic is: if it looks like a star, and quacks like a star, ... 

\section*{Acknowledgements}
We thank Tomer Shenar, Jim Fuller, Eliot Quataert, Todd Thompson, Kento Masuda, Megan Bedell, and Katelyn Breivik for useful discussions. This research was supported in part by the National Science Foundation under Grant No. NSF PHY-1748958. Hans-Walter Rix and Rhys Seeburger acknowledge support for this work by the GIF grant GIF P.S.ASTR1475.

%%%%%%%%%%%%%%%%%%%%%%%%%%%%%%%%%%%%%%%%%%%%%%%%%%
\section*{Data Availability}
Data used in this study are available upon request from the corresponding author. 



\bibliographystyle{mnras}

\appendix
\section{APOGEE spectra of the Giraffe}
\label{sec:apogee}
\begin{figure*}
    \centering
    \includegraphics[width=\textwidth]{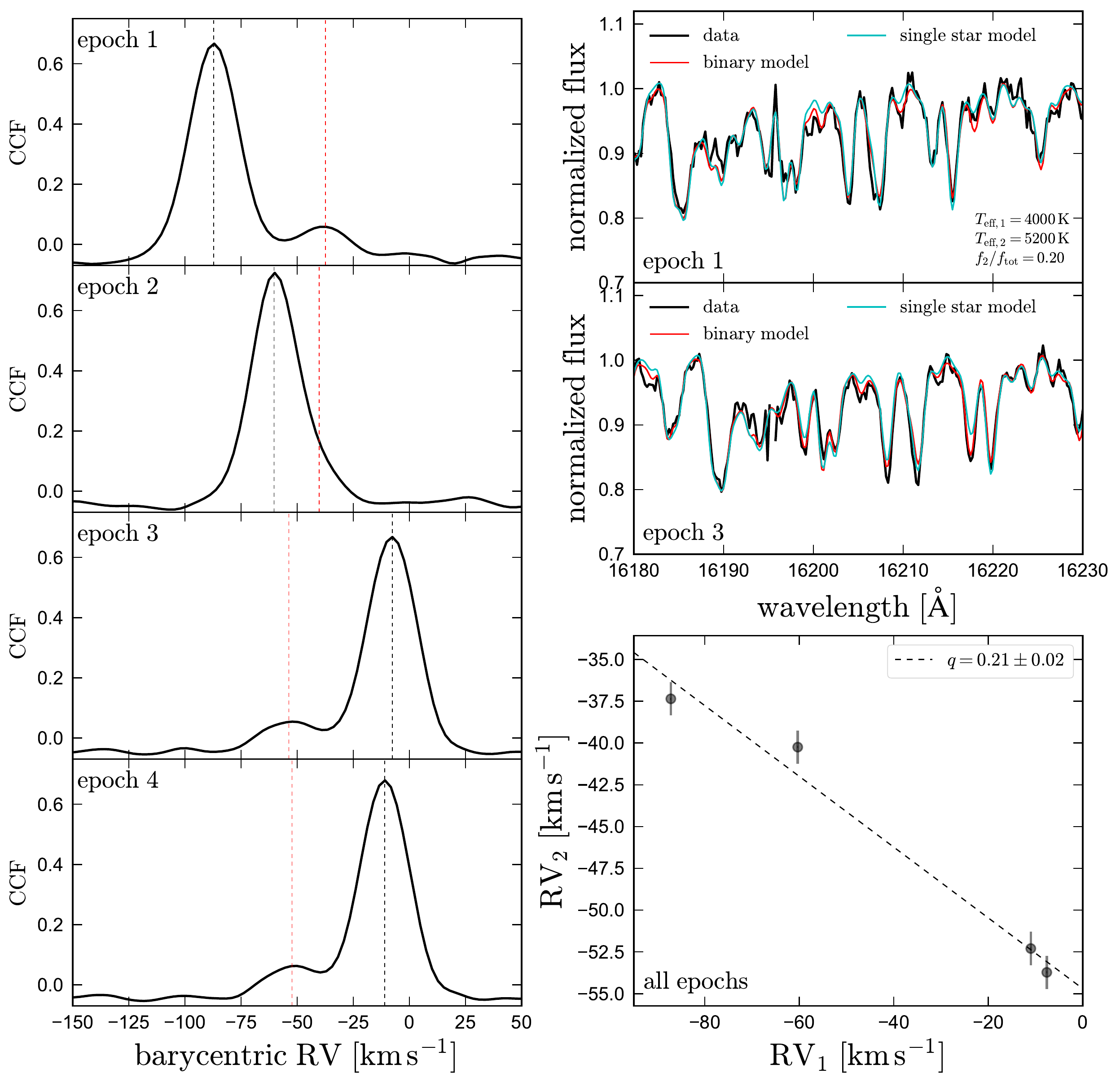}
    \caption{Detection of the Giraffe's secondary in APOGEE spectra. Left panels show cross-correlation functions (CCFs) of the individual visit spectra with a giant star template. Two CCF peaks are evident in 3 of the 4 visits, suggesting the presence of two luminous components. Dashed black and red lines show the RVs we infer for the giant and secondary by fitting the spectra with a binary model. Upper right panels show the result of fitting two visits with a binary spectral model: red line shows the best-fit binary model, and cyan shows the best-fit single-star model. The binary model provides a significantly better fit, with a total $\chi^2$ difference of about 10,000 per epoch. Clear differences in line profile shape are evident between epochs, as expected in a double-lined system. The secondary contributes 20\% of the $H-$band flux in these fits. Bottom right panel shows our inferred RVs of both components over 4 epochs. The epoch-to-epoch RV shifts of the secondary are 5 times smaller than those of the primary. This implies a mass ratio $q=M_{\rm giant}/M_2=0.21\ \pm 0.02$, consistent with what \citet{Jayasinghe2022} found with optical spectra.  }
    \label{fig:apogee_summary}
\end{figure*}

\begin{table*}
\caption{Inferred parameters from fitting APOGEE spectra of the Giraffe with a binary model. Uncertainties reflect the scatter across different visits. ${\rm RV}_i$ values are the radial velocities of both components in the 4 epochs shown in Figure~\ref{fig:apogee_summary}. We assume both components have the same [Fe/H] and [$\alpha$/Fe]. $f/f_{\rm tot}$ is the continuum flux ratio at APOGEE wavelengths, which is modeled as constant over 15,000--17,000\,\AA. }
\label{tab:apogee}
\begin{tabular}{lllllllllll}
component            & $T_{\rm eff}$ & $\log(g/{\rm cm\,s^{-2}})$ & $v_{\rm rot} \sin i$ & $\rm RV_1$         & $\rm RV_2$         & $\rm RV_3$         & $\rm RV_4$         & {[}Fe/H{]} & {[}$\alpha/ \rm Fe${]} & $f/f_{\rm tot}$ \\
\hline
                     & {[}kK{]}       &                            & $[\rm km\,s^{-1}]$   & $[\rm km\,s^{-1}]$ & $[\rm km\,s^{-1}]$ & $[\rm km\,s^{-1}]$ & $[\rm km\,s^{-1}]$ & dex        & dex               &                   \\
\hline
primary     & $4.05 \pm 0.1 $   & $1.6 \pm 0.1 $ & $11 \pm 2 $ & -87.18  & -60.32 & -7.61  & -11.00  & $-0.55 \pm 0.06$  & $0.15 \pm 0.05$   & $0.80 \pm 0.01$ \\
secondary & $5.2 \pm 0.2 $   & $3.4 \pm 0.3 $  & $16 \pm 4$ & -37.3 & -40.2  & -53.7  & -52.3  &     --    & --             &     $0.20 \pm 0.01$
\end{tabular}
\end{table*}

The Giraffe was observed 4 times by the APOGEE survey \citep{Majewski2017}; these data were also used by \citet{Jayasinghe2022}. We analyzed the DR17 version \citep{Abdurrouf2021} of the \texttt{apStar} and \texttt{apVisit} files. The results of our analysis are summarized in Figure~\ref{fig:apogee_summary} and Table~\ref{tab:apogee}.

We first inspected the cross-correlation functions (CCFs) of the individual visit spectra (left panel of Figure~\ref{fig:apogee_summary}). These are produced for all APOGEE sources as a routine part of data processing \citep{Nidever2015} by cross-correlating the individual visit spectra with a library of stellar template spectra. In DR17, the published CCFs are those produced by the best-fit single-star template spectrum.  In 3 of the 4 APOGEE visits, the CCFs contain two distinct significant peaks, as is typical of double-lined binaries \citep[e.g.][]{Kounkel2021}. This suggests that a second luminous source contributes significantly to the spectra. Both peaks are relatively narrow, implying that neither component has a large projected rotation velocity. 

Using the binary spectral fitting approach described by \citet{El-Badry2018}, we fit each of the APOGEE visits with a sum of two single-star spectral models. To predict the spectra of the component stars, we used the library of $H-$band Kurucz spectra described by \citet{Ting_2019}. These are based on 1D LTE ATLAS/SYNTHE models \citep[][]{Kurucz_1970, Kurucz_1979, Kurucz_1993}, with a linelist calibrated to the spectra of the Sun and Arcturus. We interpolate between spectral models using a neural network, as described by \citet{El-Badry2018_theory} and \citet{Ting_2019}. Observed and synthetic spectra are pseudo-continuum normalized using a Chebyshev polynomial fit to regions of the spectrum without strong lines, which were identified by \citet{Ness2015}.

The free parameters of the fit are the effective temperature ($T_{\rm eff}$), surface gravity ($\log g$), projected rotation velocity ($v\,\sin i$), and radial velocity (RV) of both stars, the $H-$band flux ratio ($f/f_{\rm tot}$), and the metallicity [Fe/H] and $\alpha$- abundance [$\alpha$/Fe]. We assume that the abundance of all $\alpha$ elements is varied in lockstep, and that all other metals trace Fe according to the solar abundance pattern from \citet{Asplund2009}. We also assume the two stars have the same surface abundances.  In contrast to \citet{El-Badry2018}, we do not require both stars to fall on a single isochrone, since we suspect the components' evolution to differ from single-star evolutionary tracks due to their mutual interaction.

Our inferred parameters are listed in Table~\ref{tab:apogee}. 
To infer uncertainties on spectral parameters that are more realistic than the formal fitting errors, we fit the 4 visits separately, reporting the mean value and standard deviation across visits. Small cutouts of fits to two of the single-epoch spectra are shown in Figure~\ref{fig:apogee_summary}, where we also plot the inferred RVs for both components. 

There is little doubt that a second component contributes significantly to the observed spectra: there are many absorption lines that are obviously better fit with a binary model. The total $\chi^2$ difference between the best-fit single star and binary models across all visit spectra is 30,000. The difference is even larger, increasing to 55,000, if we fit all 4 visit simultaneously; i.e., requiring that the same star or pair of stars can explain all of them. This is because the presence of the second component changes the morphology of the composite spectrum from epoch to epoch. Compare, for example, the relative depth of the pair of lines at $\approx 16220$\,\AA\,in Figure~\ref{fig:apogee_summary}: the lines have almost equal depth in epoch 3, but the line on the left is significantly shallower in epoch 1. 

We find that the RVs of the two components vary in anti-phase with each other. The ratio of their epoch-to-epoch RV shifts allows us to measure a dynamical mass ratio of $M_{\rm giant}/M_2 = 0.21\pm 0.02$. This is perfectly consistent with the mass ratio found by \citet{Jayasinghe2022} in the optical, and strongly implies that secondary we detect in the APOGEE spectra is the same one detected in the optical.

This fitting also allows us to measure the flux ratio in the $H$-band, which serves to constrain the radius of the secondary. We find $f_2/f_{\rm tot}$ between 0.19 and 0.22 in all visits and adopt $f_2/f_{\rm tot} =0.2 \pm 0.01$ as the fiducial value. This is larger than the flux contribution inferred by \citet{Jayasinghe2022} by extrapolating from the optical. However, it is likely that their analysis underestimated the contributions of the secondary as discussed in Section~\ref{sec:radius}.

Our fitting yields an effective temperature of $4050\pm 100$\,K for the primary, and $5200\pm 200$ for the secondary. We find slow rotation values of $11\pm 2\,\rm km\,s^{-1}$ and $16\pm 4\rm km\,s^{-1}$ for the primary (the giant) and the secondary (the subgiant). The inferred metallicity is subsolar, $\rm [Fe/H] = -0.55\pm 0.06$. The temperature we infer for the giant when fitting a binary model is 100\,K cooler than the value we infer with a single-star model.  Because the subgiant is warmer than the giant (though still cooler than the value inferred by \citealt{Jayasinghe2022}, which was $T_{\rm eff} = 6350\pm 700$\,K), its fractional contribution to the total spectrum is expected to increase toward bluer wavelengths. This is consistent with our findings from the optical spectra (Figure~\ref{sec:HIRES}).

% Don't change these lines
\bsp	% typesetting comment
\label{lastpage}
\end{document}